\documentclass[sigconf]{acmart}
\usepackage{enumitem}
\usepackage{bbm}
\usepackage{multirow}
\usepackage{caption}
\usepackage{graphicx}
\usepackage{float} 
\usepackage{subcaption}
\usepackage{colortbl}
\usepackage{xcolor}
\usepackage{balance}
\usepackage{multicol}
\usepackage[linesnumbered, ruled]{algorithm2e}
\SetKwRepeat{Do}{do}{while}%
\makeatletter
\DeclareRobustCommand\onedot{\futurelet\@let@token\@onedot}
\def\@onedot{\ifx\@let@token.\else.\null\fi\xspace}

\definecolor{myPink}{RGB}{255, 217, 224}

\def\eg{\emph{e.g}\onedot} 
\def\ie{\emph{i.e}\onedot}

 \def\model{\textsc{SEAN}}
\def\BS{Representative Neighbor Selector} \def\DS{Temporal-aware Aggregator}
\makeatother

\AtBeginDocument{%
  \providecommand\BibTeX{{%
    \normalfont B\kern-0.5em{\scshape i\kern-0.25em b}\kern-0.8em\TeX}}}

\setcopyright{acmcopyright}
\copyrightyear{2024}
\acmYear{2024}
\setcopyright{acmlicensed}\acmConference[KDD '24]{Proceedings of the 30th
ACM SIGKDD Conference on Knowledge Discovery and Data Mining}{August
25--29, 2024}{Barcelona, Spain}
\acmBooktitle{Proceedings of the 30th ACM SIGKDD Conference on Knowledge
Discovery and Data Mining (KDD '24), August 25--29, 2024, Barcelona, Spain}
\acmDOI{10.1145/3637528.3671877}
\acmISBN{979-8-4007-0490-1/24/08}

\begin{document}

\title{Towards Adaptive Neighborhood for Advancing Temporal Interaction Graph Modeling}

\author{Siwei Zhang}
  \email{swzhang22@m.fudan.edu.cn}
  \affiliation{%
    \institution{Shanghai Key Laboratory of Data Science, School of Computer Science, Fudan University}
    \city{Shanghai}
    \country{China}
   }

\author{Xi Chen}
\author{Yun Xiong}
\authornote{Corresponding author}
  \email{x_chen21@m.fudan.edu.cn}
   \email{yunx@fudan.edu.cn}
  \affiliation{%
    \institution{Shanghai Key Laboratory of Data Science, School of Computer Science, Fudan University}
    \city{Shanghai}
    \country{China}
  }

\author{Xixi Wu}
  \email{21210240043@m.fudan.edu.cn}
  \affiliation{%
    \institution{Shanghai Key Laboratory of Data Science, School of Computer Science, Fudan University}
    \city{Shanghai}
    \country{China}
  }

\author{Yao Zhang}
\author{Yongrui Fu}
  \email{yaozhang@fudan.edu.cn}
  \email{23210240154@m.fudan.edu.cn}
  \affiliation{%
     \institution{Shanghai Key Laboratory of Data Science, School of Computer Science, Fudan University}
     \city{Shanghai}
      \country{China}
  }

\author{Yinglong Zhao}
  \email{zhaoyinglong.zyl@antgroup.com}
  \affiliation{%
     \institution{Ant Group}
     \city{Shanghai}
      \country{China}
  }

\author{Jiawei Zhang}
  \email{jiawei@ifmlab.org}
  \affiliation{%
     \institution{IFM Lab, Department of Computer Science, University of California, Davis}
     \state{CA}
     \country{USA}
  }

\renewcommand{\shortauthors}{Siwei Zhang, et al.}

\begin{abstract}
Temporal Graph Networks (TGNs) have demonstrated their remarkable performance in modeling temporal interaction graphs. These works can generate temporal node representations by encoding the surrounding neighborhoods for the target node. However, an inherent limitation of existing TGNs is their reliance on \textit{fixed}, hand-crafted rules for neighborhood encoding, overlooking the necessity for an adaptive and learnable neighborhood that can accommodate both personalization and temporal evolution across different timestamps. In this paper, we aim to enhance existing TGNs by introducing an \textit{adaptive} neighborhood encoding mechanism. We present \textbf{{\model}} (\underline{\textbf{S}}elective \underline{\textbf{E}}ncoding for \underline{\textbf{A}}daptive \underline{\textbf{N}}eighborhood), a flexible plug-and-play model that can be seamlessly integrated with existing TGNs, effectively boosting their performance. To achieve this, we decompose the adaptive neighborhood encoding process into two phases: (i) representative neighbor selection, and (ii) temporal-aware neighborhood information aggregation. Specifically, we propose the Representative Neighbor Selector component, which automatically pinpoints the most important neighbors for the target node. It offers a tailored understanding of each node's unique surrounding context, facilitating personalization. Subsequently, we propose a Temporal-aware Aggregator, which synthesizes neighborhood aggregation by selectively determining the utilization of aggregation routes and decaying the outdated information, allowing our model to adaptively leverage both the contextually significant and current information during aggregation. We conduct extensive experiments by integrating {\model} into three representative TGNs, evaluating their performance on four public datasets and one financial benchmark dataset introduced in this paper. The results demonstrate that {\model} consistently leads to performance improvements across all models, achieving SOTA performance and exceptional robustness.

\end{abstract}

\ccsdesc[500]{Information systems~Data mining}
\ccsdesc[500]{Computing methodologies~Learning latent representations}
\ccsdesc[300]{Computing methodologies~Neural networks}

\keywords{Temporal Graph Networks; Representation Learning; Data Mining}


\maketitle

\section{Introduction}
\label{sec:intro}
\begin{figure}
    \centering
    \includegraphics[width=\linewidth]{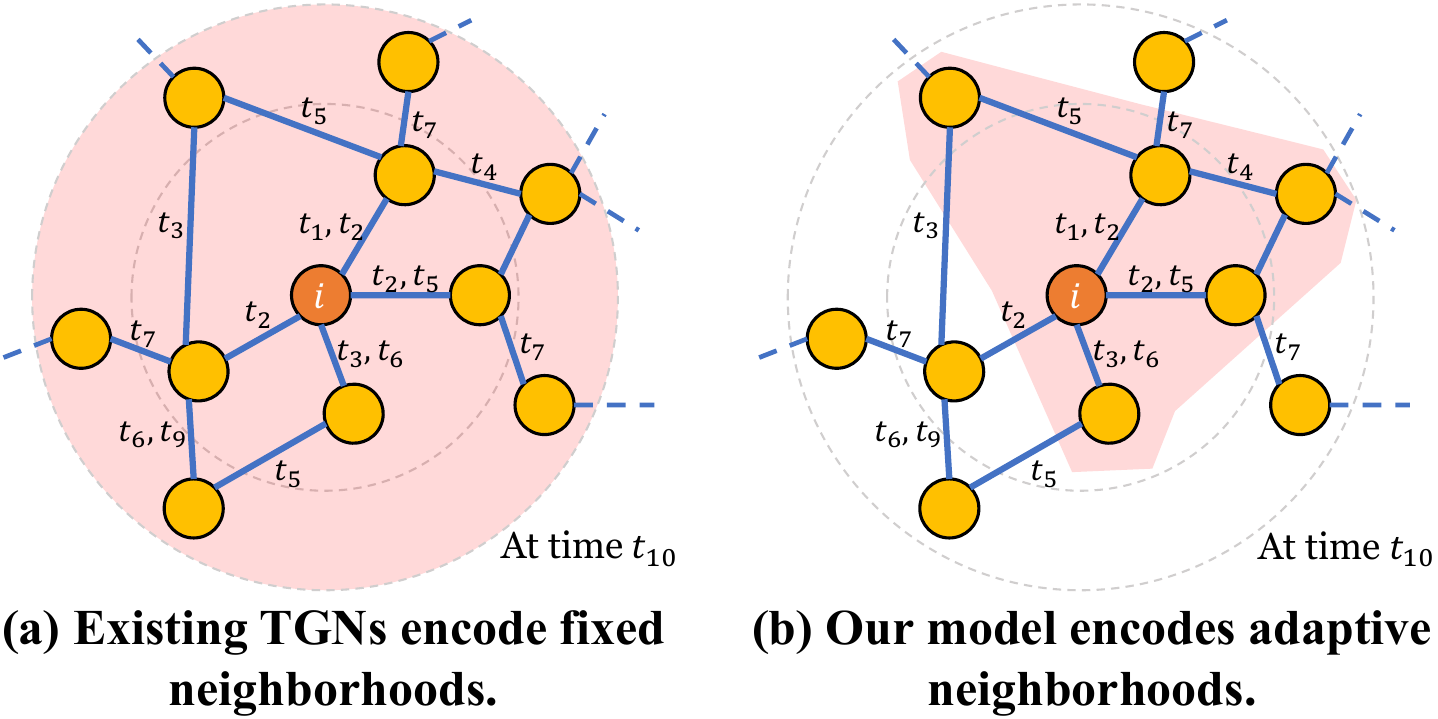}
    \caption{Comparison between fixed and adaptive neighborhoods during the encoding process. (a) Existing TGNs \cite{tgn,tiger,dygformer} always adopt fixed rules for neighborhood encoding, \eg, encode 2-hop neighborhoods. (b) We propose adaptive neighborhood encoding to facilitate both the personalized and temporal understanding of a target node.}
    \label{fig:introduction}
\end{figure}
Temporal Interaction Graphs (TIGs) can model the dynamic graph-structured data in many real-world application scenarios, where objects are depicted as nodes and timestamped interactions between them are represented as edges \cite{tiger}. Unlike static graphs, TIGs exhibit dynamic changes over time. To effectively capture the dynamic nature of TIGs and facilitate representation learning, extensive research has been conducted on developing Temporal Graph Networks (TGNs) \cite{dyrep, tgat, tgn, tiger}. These TGNs can generate temporal node representations by encoding the neighborhoods around the target node, thus enabling downstream predictions.


Despite the remarkable success of existing TGNs, a fundamental weakness inherent in their designs is the reliance on the \textbf{fixed}, hand-crafted rules for neighborhood encoding, failing to account for the necessary personalization and the temporal evolution across different timestamps. For example, when generating the representation for a target node at a specific timestamp, as shown in Figure~\ref{fig:introduction}a, existing TGNs, like TIGE \cite{tiger}, indiscriminately encode the current 2-hop neighborhood. Different from such existing works, we emphasize that an \textbf{adaptive} neighborhood encoding mechanism is crucial for generating more expressive representations.

\textbf{Personalization.} 
Personalization is essential for TIG representation learning at both graph-scale and node-scale levels. Different TIGs exhibit distinct characteristics \cite{edgebank}, and a universal, pre-defined rule for neighborhood encoding is inflexible \cite{messagepassing}. For example, while the $k$-hop assumption may work well for low-density TIGs, it could lead to indistinguishable representations in denser ones \cite{dense}. Additionally, even within a single TIG, the suitable neighborhoods for nodes can differ significantly. This hypothesis is reasonable due to the diverse neighborhood connectivity patterns and potential noise. For instance, in financial networks, transaction patterns of banks arise from different factors \cite{finacialimpact, finacialimpact2}, and often contain noise irrelevant to their primary financial interests \cite{rdgsl, noise}.

\textbf{Temporal evolution.}
A suitable neighborhood for a target node should adapt to different timestamps to align with the temporal evolution of TIGs, necessitating a temporal-aware design for neighborhood encoding. For example, investors' preferences may change according to economic cycles, such as preferring technology stocks during technological booms while consumer staples during economic downturns. Fixed rules for neighborhood encoding cannot accommodate to such preference shifts in the temporal dimension, leading to suboptimal representations for effectively capturing these changing preferences.

Based on the aforementioned motivations, as depicted in Figure~\ref{fig:introduction}b, it becomes critical to allow for a more flexible, scalable, and robust algorithm that performs adaptive neighborhood encoding obeying both personalization and temporal awareness.

In this paper, we aim to enhance existing TGNs by adaptively encoding personalized and temporal-aware neighborhoods through the introduction of a convenient plug-and-play model, referred to as \textbf{{\model}} (\underline{\textbf{S}}elective \underline{\textbf{E}}ncoding for \underline{\textbf{A}}daptive \underline{\textbf{N}}eighborhood). Our {\model} can significantly boost existing TGNs, and it comprises two main components:
(i) {\BS}.
To effectively select personalized neighborhoods for nodes within TIGs, we introduce {\BS}. It refines the neighborhood by choosing representative neighbors who are important to the target node. However, an overemphasis on these neighbors can result in a homogeneous neighborhood that lacks the necessary diversity~\cite{classincrease}. Therefore, we incorporate a penalty mechanism that penalizes the over-concentration of neighbors, maintaining a balanced and personalized neighborhood for the target node.
(ii) {\DS}.
To achieve temporal-aware neighborhood encoding, we further propose the {\DS} that automatically aggregates neighborhoods with the temporal understanding of the target node. Specifically, it takes charge of neighborhood aggregation by explicitly determining the utilization of each aggregation route, facilitating the adaptive route aggregation or pruning as needed. Meanwhile, our outdated-decay mechanism strategically de-emphasizes those outdated routes, allowing our model to concentrate on more fresh and up-to-date information. 

In summary, our main contributions are:
\begin{itemize}[itemsep=3pt, parsep=3pt, leftmargin=*]
\item We focus on adaptive neighborhood encoding for temporal interaction graph (TIG) modeling and propose {\model}. {\model} is the first model proposed to automate neighborhood encoding in TIG modeling, and it can be easily adopted to boost existing TGNs.
\item We develop a {\BS}, which enables the model to effectively choose representative neighbors for the target node, achieving personalized neighborhood encoding.
\item We propose a {\DS}, which aggregates neighborhoods by explicitly determining the utilization of aggregation routes and decaying the outdated information, facilitating temporal-aware neighborhood encoding.
\item We conduct extensive experiments on several TIG benchmark datasets to validate {\model}'s effectiveness. Furthermore, we introduce \textbf{TemFin}, a new financial transaction TIG benchmark dataset in this paper. This benchmark represents a more challenging environment due to its complexity in the financial domain.
\end{itemize}



\section{Related Work}
\subsection{Temporal Graph Networks (TGNs)}
Temporal Graph Networks (TGNs) are designed to generate temporal node representations of TIGs by encoding the neighborhoods for the target node at any given timestamp \cite{cope, rdgsl, ilore, speed}. They typically encode their neighborhoods based on a fixed, pre-defined rule. According to how these neighborhoods are pre-defined, existing TGNs can be categorized into two types: Breadth-First Search TGNs (BFS-TGNs) and Depth-First Search TGNs (DFS-TGNs). 

BFS-TGNs \cite{jodie, dyrep, tgat, tgn, tiger, rdgsl, ilore, promptTIG, magic} prioritize selecting neighborhoods close to the target node. They then encode these neighborhoods via various aggregation functions, generating representations for the target node. Different from BFS-TGNs, DFS-TGNs \cite{pint, caw} utilize some temporal walks starting from the target node as their neighborhoods, and encode these walks through the frequency of node occurrences. Currently, DyGFormer \cite{dygformer}, one of BFS-TGNs, encodes the first-order neighborhoods through Transformer blocks, achieving SOTA results among existing TGNs.

Different from existing TGNs, our model goes beyond fixed rules for neighborhood encoding. It adaptively encodes the neighborhoods that contribute most to the representations, thus enhancing the model's overall effectiveness.

\subsection{Adaptive Neighborhood for Graph Learning}
Graph learning with adaptive neighborhoods aims to encode undetected or enhanced neighborhoods for better representation learning, which has achieved remarkable performance on static graphs. 

Early works \cite{graphnas, autognn, policygnn, hypernetworks} encode adaptive neighborhoods by utilizing Reinforcement Learning (RL). However, these works face significant limitations due to the immense search space and the non-differentiability of models, resulting in substantial computational overhead. To address this issue, GeniePath \cite{geniepath} employs a differentiable module for adaptive neighborhood encoding, reducing the model complexity and achieving better performance. Subsequently, an increasing number of adaptive neighborhood encoding methods \cite{jump, same, subgraph, messagepassing, pooling, autogel, beyond, luo} have been proposed, significantly empowering the representation learning on static graphs.

These existing methods cannot be directly adopted in TIGs, primarily because TIGs exhibit significant temporal differences compared to static graphs. For instance, TIGs are consistently characterized by a sequence of timestamped interactions, rather than by adjacency matrices that are typical in static graphs, due to the dynamic nature of TIGs.






\section{Preliminaries}
\label{sec:preliminaries}
\begin{figure*}
    \centering
    \includegraphics[width=\linewidth]{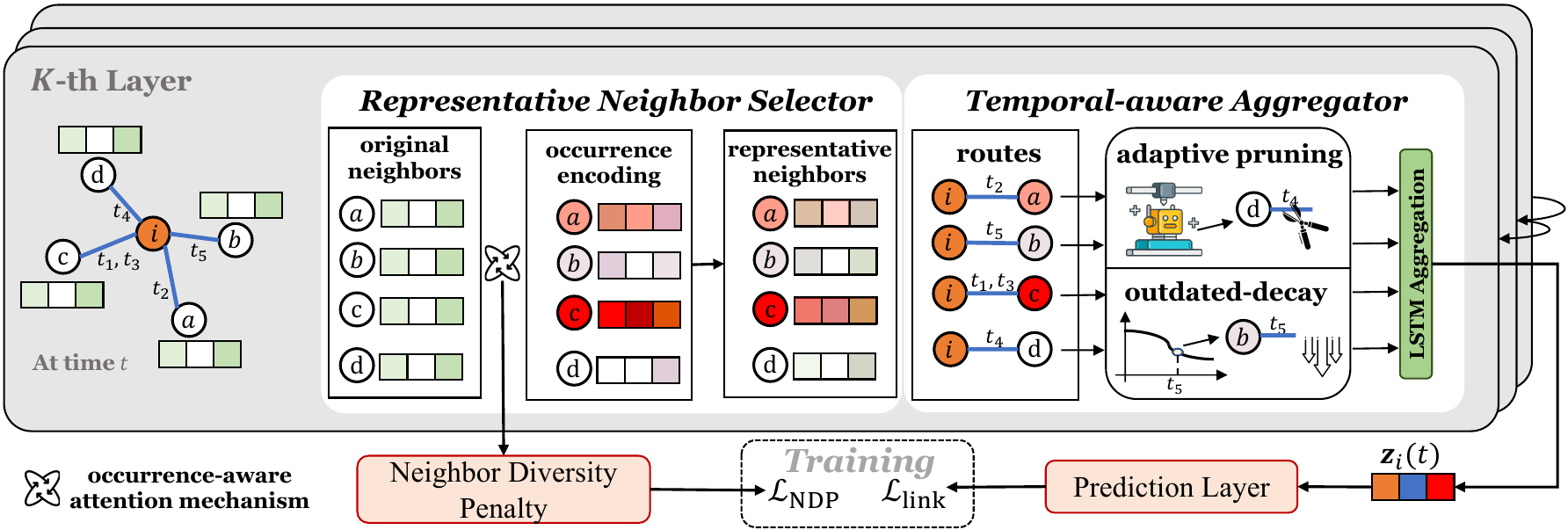}
    \caption{Framework of the proposed plug-and-play model. Our {\BS} can empower the model to pinpoint the important neighbors, who then act as the personalized representatives for the target node. Meanwhile, we propose the neighbor diversity penalty to penalize the over-concentration of these neighbors, thus maintaining a more balanced neighborhood. Furthermore, we conduct our Temporal-aware Aggregator by LSTM aggregation, where we propose an adaptive pruning module that explicitly determines whether to aggregate information from the given route or to prune it as needed, and an outdated-decay mechanism that de-emphasizes the outdated information. Finally, we generate the temporal node representations for downstream tasks by extracting the encoded neighborhood information from the $K$-th layer of {\model}.}
    \label{fig:method}
\end{figure*}

\subsection{Problem Formulation}

\begin{definition}
    \textbf{Temporal Interaction Graph.} Given%
    \footnote{For simplicity, in this paper, we can just represent a node by its index, \ie, an integer.} a node set $\mathcal{V} = \left\{1, ..., |\mathcal{V}| \right\}$, a temporal interaction graph can be represented as a sequence of edge set $\mathcal{E} = \left\{(i, j, t)\right\}$, where $i, j \in \mathcal{V}$ and $t > 0$. Each edge $(i, j, t)$ denotes an interaction between node $i$ and node $j$ that happened at time $t$ with the feature $\mathbf{e}_{ij}(t)$. The edge feature vector can depict various information about the interactions among nodes, \eg, interaction type. Any edge $(i, j, t) \in \mathcal{E}$ only has access to its historical data before time $t$, \ie, $\left\{(i, j, \tau) \in \mathcal{E} | \tau < t \right\}$.
\end{definition}

\begin{definition}
    \textbf{Temporal Interaction Graph Modeling.} Given an edge $(i, j, t) \in \mathcal{E}$ and its historical data $\left\{(i, j, \tau) \in \mathcal{E} | \tau < t \right\}$, temporal interaction graph modeling aims to learn a mapping function $f: (i, j, t) \mapsto \mathbf{z}_i(t), \mathbf{z}_j(t)$, where $\mathbf{z}_i(t), \mathbf{z}_j(t) \in \mathbbm{R}^d$ represent the representations of nodes $i$ and $j$ at time $t$, respectively, and $d$ is the representation vector dimension.
\end{definition}

\subsection{Temporal Embedding Module}\label{sec:temporalembeddingmodule}
As a crucial component within most existing TGNs, temporal embedding module \cite{tiger} is responsible for generating temporal representations for any given node $i$ at time $t$, $\mathbf{z}_i(t)$, which involves a $K$-layer temporal attention network to encode $i$'s current neighborhood. In this section, we generalize various temporal embedding modules employed in different TGNs into a cohesive framework. This unification allows us to provide a comprehensive overview of the neighborhood encoding process within TGNs.


For the target node $i$ at time $t$ of the $k$-th layer, $k \in \{1, ..., K\}$, temporal embedding module aggregates $i$'s neighborhood information $\widetilde{\mathbf{h}}_i^{(k)}(t)$ and $i$'s representation from the previous layer $\mathbf{h}_i^{(k-1)}(t)$ using a Multi-Layer Perceptron (MLP) as the aggregator:
\begin{equation}\label{eq:aggregation}
    \mathbf{h}_i^{(k)}(t) = \text{MLP}^{(k)} \left( \mathbf{h}_i^{(k-1)}(t) \| \; \widetilde{\mathbf{h}}_i^{(k)}(t)\right),
\end{equation}
where $\|$ denotes the concatenation operation. The neighborhood information $\widetilde{\mathbf{h}}_i^{(k)}$ is obtained through an attention mechanism \cite{tgat}, which involves assigning attention scores to $i$'s neighborhood:
\begin{equation}\label{eq:attentionscore}
    \widetilde{\mathbf{h}}_i^{(k)}(t) = \text{Softmax}\left( \mathbf{a}_i^{(k)}(t)\right) \cdot \mathbf{V}_{i}^{(k)}(t), 
\end{equation}
where $\mathbf{a}_i^{(k)}(t) = [a_{ij}^{(k)}(t)]_{j \in \mathcal{N}_i(t)}$ represents the attention vector of node $i$, and the matrix $\mathbf{V}_{i}^{(k)}(t) = [\mathbf{v}_{ij}^{(k)}(t)]_{j \in \mathcal{N}_i(t)}$ encapsulates the messages from $i$'s neighborhood. Each element $a_{ij}^{(k)}(t)$ is the attention score from node $i$ to its neighboring node $j \in \mathcal{N}_i(t)$, which is computed as follows: 
\begin{equation}\label{eq:score_i_j}
    a_{ij}^{(k)}(t) = \frac{f_{\text{q}}\left( \mathbf{h}_i^{(k-1)}(t) \right) f_{\text{k}} \left(\mathbf{h}_j^{(k-1)}(t)\right) ^T}{\sqrt{d^{(k)}}},
\end{equation}
and each row $\mathbf{v}_{ij}^{(k)}(t)$ is the message carried from neighbor $j$, which is determined by $\mathbf{v}_{ij}^{(k)}(t) = f_{\text{v}}(\mathbf{h}_j^{(k-1)}(t))$. In the above equations, $f_{*}(\cdot) (* \in \{\text{q, k, v}\})$ represents the encoding functions for queries, keys, and values, respectively \cite{transformer, graformer}. These functions may have different specific representations for different TGNs \cite{tgat,tgn,tiger}. For simplicity, we consider a single-head attention formula in this paper. By incorporating personalized and temporal-aware designs into the neighborhood attention mechanism (Equation~\ref{eq:score_i_j}) and the aggregation process (Equation~\ref{eq:aggregation}), we can elevate this module to achieve adaptive neighborhood encoding. This innovative design enables our model to serve as a seamlessly integratable plug-and-play enhancement for various existing TGNs, broadening their flexibility and effectiveness.

\section{Methodology}
As we have mentioned, adaptive neighborhood encoding requires both personalized and temporal considerations. Therefore, we decompose this process into two phases, \ie, a {\BS} first assigns distinctive attention to select important neighbors for personalization, and our {\DS} then aggregates the neighborhood information to ensure temporal relevance. We will introduce these components in the following subsections.

\subsection{\BS}\label{sec:bs}
We propose the occurrence-aware attention and neighbor diversity penalty mechanisms to select personalized representative neighborhoods for the target node.

\subsubsection{Occurrence-aware Attention Mechanism}
In Equation \ref{eq:score_i_j}, the attention score is calculated based on the semantic correlation between the target node and its neighboring nodes. However, this semantics-based attention computation may be suboptimal to select representative neighborhoods due to its lack of personalization and adaptability \cite{tgat, graformer}. Here, we propose the occurrence-aware attention mechanism to enhance the traditional semantics-based attention mechanism for selecting important neighbors as personalized representatives. Intuitively, neighbor occurrence, which quantifies how frequently a neighbor occurs, is a significant signal in identifying the importance of each neighbor for a given target node. Frequently occurring neighbors are more likely to be important and possess more relevant information, thus making them more effective and convincing representatives.

Formally, for node $i$ at time $t$ and its historical neighbors $\mathcal{N}_i(t)$, we first count how often each neighbor has occurred historically and derive it to a one-dimensional occurrence frequency feature, which is represented by $\mathbf{f}_i(t) \in \mathbb{R}^{ |\mathcal{N}_i(t)| \times 1}$. To balance the raw counts, we then normalize this feature by dividing its temporal node degree $\mathbf{d}_i{(t)} \in \mathbb{R}^{ |\mathcal{N}_i(t)| \times 1}$, which is denoted as:
\begin{equation}
    \hat{\mathbf{f}}_i(t) = \mathbf{f}_i(t) \odot  \mathbf{d}_i^{-1}(t) \in \mathbb{R}^{ |\mathcal{N}_i(t)| \times 1},
\end{equation}
where $\odot$ denotes the element-wise product operation. The normalized occurrence frequency feature vector both considers the interaction frequency and mitigates the unexpected adverse impacts from the hub nodes (\ie, those with large degrees), which are known as ``degree-related biases'' \cite{normalize1, normalized2}. To enhance the expressiveness of the occurrence information, we apply a function $f_{\text{e}}(\cdot)$ to encode our normalized occurrence frequency feature $\hat{\mathbf{f}}_i(t)$ into a high-dimensional occurrence encoding as follows:
\begin{equation}
    \mathbf{R}_i(t) = f_{\text{e}}\left(\hat{\mathbf{f}}_i(t) \right) \in \mathbb{R}^{ |\mathcal{N}_i(t)| \times d}.
\end{equation}
Similar to DyGFormer \cite{dygformer}, we implement $f_{\text{e}}(\cdot)$ by a two-layer MLP whose input and output dimensions are 1 and $d$, respectively.

To further enrich the occurrence-aware attention computation, we incorporate both edge attributes and temporal information. For each neighbor node $j \in \mathcal{N}_i(t)$, we retrieve its occurrence encoding as $\mathbf{r}_{ij}(t) = \mathbf{R}_{i, j:}(t) \in \mathbb{R}^{d}$. Our occurrence-aware attention score from node $i$ to node $j$ at time $t$, $\tilde{a}_{ij}(t)$, can be represented by:
\begin{equation}\label{eq:occurrence-basedscore}
    \tilde{a}_{ij}(t) = \text{tanh}\left( \mathbf{w}_{i} \cdot [ \mathbf{r}_{ij}(t) \| \mathbf{e}_{ij}(t) \| \phi(t - t_{j})  ]  \right),
\end{equation}
where $\phi(\cdot)$ is the commonly used time encoding function in TGNs \cite{tiger, tgat}, and $\mathbf{w}_i \in \mathbb{R}^{1 \times 3d}$ represents the trainable reshaping vector.

Finally, we update the traditional attention score $a_{ij}(t)$ with our new occurrence-aware attention score $\tilde{a}_{ij}(t)$ to obtain the final attention score as $\hat{a}_{ij}(t) = a_{ij}(t) + \tilde{a}_{ij}(t)$. This combined score integrates both semantic relevance and occurrence frequency, enabling our model to select important neighbors as representatives.

\subsubsection{Neighbor Diversity Penalty}\label{sec:NDP}
Although our occurrence-aware attention greatly helps the model in identifying the most important neighbors as the representatives for the target node, an overemphasis on these neighbors can result in a homogeneous neighborhood that lacks the necessary diversity. For example, in the world of finance, an abundance of homogeneous investment choices may lead to the undesirable phenomenon known as the ``Financial Echo Chamber'' \cite{finacialecho}, analogous to the ``Information Cocoon'' in Recommender Systems \cite{informationcoco, informationcoco1}. To address this issue, we propose an additional neighbor diversity penalty mechanism to encourage the selection of more diverse neighbors. By leveraging this mechanism, our approach aims to strike a balance between prioritizing important neighbors and ensuring their diversity, thereby enhancing the overall effectiveness of our model.

Let $\mathbb{I}(\cdot)$ be the indication function \cite{attentiondiversity}, and $\hat{a}_{ij}^{(k)}$ represents the occurrence-aware attention score from node $i$ to its neighbor $j \in \mathcal{N}_i(t)$ in the $k$-th layer%
\footnote{For clarity, we omit the timestamp term $t$ in the remaining part of our paper, unless specified otherwise.}. To ensure that our model penalizes the over-important neighbors with excessively high attention scores, we first employ a score-filtering technique in the attention scores based on a threshold $\tau \in \left[0, 1\right]$, beyond which the scores are preserved. We have:
\begin{equation}\label{indication}
\hat{a}_{ij}^{(k)}=\mathbb{I}\left(\hat{a}_{ij}^{(k)} \geq \tau \right) \cdot \hat{a}_{ij}^{(k)}.
\end{equation}
Subsequently, we calculate the average score within $i$'s neighbors by $\bar{a}^{(k)}_i := \frac{1}{N}\sum_{j}\hat{a}_{ij}^{(k)}$ where $N = |\mathcal{N}_i(t)|$ is the total number of neighbors. Then, we compute the product similarity between $i$'s neighbors' attention scores and their corresponding average score by $s^{(k)}_i = \frac{1}{N}\sum_{j}\hat{a}_{ij}^{(k)}\cdot\bar{a}^{(k)}_{i}$. Finally, we minimize the dissimilarity across nodes and layers, promoting alignment and coherence within our attention mechanism as follows:
\begin{equation}\label{NDPloss}
    \mathcal{L}_{\mathrm{NDP}}=-\frac{1}{K} \sum_{k=1}^{K} \mathcal{L}_{\mathrm{NDP}}^{(k)}, \quad \mathcal{L}_{\mathrm{NDP}}^{(k)}=\frac{1}{|\mathcal{V}|} \sum_{i \in \mathcal{V}} s^{(k)}_i.
\end{equation}
$\mathcal{L}_{\mathrm{NDP}}$ will serve as a controllable auxiliary loss in our model. This penalty method helps our model to prevent the over-concentration of just a few neighbors, ensuring a balanced neighborhood.

\subsection{\DS}\label{sec:DS}
To incorporate temporal understanding, we introduce a Temporal-aware Aggregator to replace the initial aggregator $\text{MLP}^{(k)}(\cdot)$ in Equation \ref{eq:aggregation}. Specifically, we implement this module using LSTM cells \cite{LSTM}, chosen for its well-known strengths in temporal awareness:
\begin{equation}
    \mathbf{h}^{(k)} = \text{AGG}^{(k)} \left(\mathbf{h}^{(k-1)}\right) = \text{LSTM}^{(k)} \left( \mathbf{h}^{(k-1)}, \widetilde{\mathbf{h}}^{(k)}, \mathbf{c}^{(k)}  \right).
\end{equation}
Here, we fetch the hidden state from the $k$-th LSTM cell $\mathbf{h}^{(k)}$ as our \textbf{output} representation of the $k$-th aggregation layer $\text{AGG}^{(k)}(\cdot)$. The \textbf{inputs} are three folds: the output representation (\ie, hidden state) from the previous layer $\mathbf{h}^{(k-1)}$, the refined neighborhood information $\widetilde{\mathbf{h}}^{(k)}$ that incorporates our proposed attention mechanism, and the cell state from the $k$-th LSTM $\mathbf{c}^{(k)}$. To fully achieve temporal-aware encoding, we enhance our LSTM aggregator through an adaptive pruning module to facilitate selective aggregates of $\mathbf{h}^{(k)}$, and an outdated-decay module to enhance the temporal sensitivity of the cell state $\mathbf{c}^{(k)}$. Note that we simplify our notation by omitting the subscript term $i$ and the time $t$. In practice, for node $i$ at time $t$, we determine its temporal representation by extracting the output representation from the last layer, \ie, $\textbf{z}_i(t) = \mathbf{h}_i^{(K)}(t)$.

\subsubsection{Adaptive Pruning Module.} \label{sec:prune}
This module not only encourages our model to focus on incorporating more crucial information for performance improvement (Section \ref{sec:ablation}) but also enhances the model interpretability, which explicitly specifies the model's decision-making process in seeking adaptive neighborhoods (Section \ref{sec:case}).

Formally, we introduce a layer state $\tilde{u}^{(k)} \in [0, 1]$ for each $k$-th layer, which represents the probability that the current route will either be aggregated or pruned. We have:
\begin{equation}\label{eq:bernulli}
u^{(k)}=f_{\text{round}}\left(\tilde{u}^{(k)}\right),
\end{equation}
where $f_{\text{round}}(\cdot)$ denotes the binarization function derived from the rounding operation and $u^{(k)} \in \{0, 1\}$. Then, $u^{(k)}$ is utilized to determine whether the current route information will be aggregated ($u^{(k)}=1$) or copied from the previous layer ($u^{(k)}=0$). We have:
\begin{equation}\label{eq:update}
    \mathbf{h}^{(k)}=u^{(k)} \cdot \mathbf{h}^{(k)}+\left(1-u^{(k)}\right) \cdot \mathbf{h}^{(k-1)},
\end{equation}
Afterward, we update the following layer state $\tilde{u}^{(k+1)}$ using the aggregated/copied information from the $k$-th layer $\mathbf{h}^{(k)}$ as follows:
\begin{align}
& \Delta \tilde{u}^{(k)}=\sigma\left(\mathbf{W}_p \cdot \mathbf{h}^{(k)}+\mathbf{b}_p\right), \\
 \tilde{u}^{(k+1)}= u^{(k)} \cdot \Delta \tilde{u}&^{(k)} + \left(1-u^{(k)}\right) \cdot \left(\tilde{u}^{(k)}+\max \left(\Delta \tilde{u}^{(k)}, 1- \tilde{u}^{(k)}\right)\right),
 \label{eq:stateupdate}
\end{align} 
where $\mathbf{W}_p \in \mathbb{R}^{d \times d}$ is the learnable matrix, $\mathbf{b}_p$ is the learnable vector, and $\sigma(\cdot)$ is the sigmoid function. During aggregation, the aggregated neighborhood progressively becomes closer to the central node as the number of aggregation layers increases. It means that higher aggregation layers aggregate more central neighborhoods that have been proven to hold more vital information \cite{dygformer}. This formulation encodes the observation that the likelihood of aggregation increases with the number of aggregation layers. Whenever the $k$-th layer aggregation is omitted, the pre-activation of the layer state for the following layer $\tilde{u}^{(k+1)}$ is incremented by $\Delta \tilde{u}^{(k)}$. Conversely, if the $k$-th layer aggregation is performed, the accumulated value will be reset and $\tilde{u}^{(k+1)} = \Delta \tilde{u}^{(k)}$. In this way, our model can explicitly determine the route aggregation or pruning as needed.

\subsubsection{Outdated-decay Module.} \label{timedecay}
Outdated-decay module aims to de-emphasize those outdated aggregation routes, ensuring the incorporation of more fresh and up-to-date information.

We believe that the long-term cell state retains important and more enduring information, while the short-term cell state is utilized for processing immediate information. Therefore, we first decompose the $k$-th layer cell sate $\mathbf{c}^{(k)}$ into two elements, \ie, the short-term cell state $\mathbf{c}^{(k)}_s$ and the long-term cell state $\mathbf{c}^{(k)}_l$. We have:
\begin{align}\label{eq:outdecay1}
    \mathbf{c}^{(k)}_s & = \tanh \left(\mathbf{W}_d \cdot \mathbf{c}^{(k)}+\mathbf{b}_d\right), \\
    \mathbf{c}^{(k)}_l & = \mathbf{c}^{(k)}-\mathbf{c}^{(k)}_s,
\end{align}
where matrix $\mathbf{W}_d \in \mathbb{R}^{d \times d}$ and vector $\mathbf{b}_d$ are learnable parameters. The short-term cell state $\mathbf{c}^{(k)}_s$ is initially produced by a neural network with the activation function. Subsequently, the long-term cell state $\mathbf{c}^{(k)}_l$ can be distinguished from the original cell state $\mathbf{c}^{(k)}$.

To prevent from forgetting too rapidly, we keep our long-term cell state $\mathbf{c}^{(k)}_l$ untouched and forget the short-term cell state $\mathbf{c}^{(k)}_s$ according to the time interval $\Delta_t = t - t^-$ between the current time $t$ and the last interaction time of the neighbor in the corresponding route $t^-$. We apply a decay mechanism as follows:
\begin{equation}
    \hat{\mathbf{c}}^{(k)}_{s} = \mathbf{c}^{(k)}_s \cdot  g\left(\Delta_t\right).
\end{equation}
The decay function $g(\Delta_t)$ is defined by $\operatorname{exp}\left\{-2\cdot\Delta_t / t_{max}\right\}$, where $t_{max}$ represents the maximum value among all time intervals. Finally, the forgotten short-term cell state $\hat{\mathbf{c}}^{(k)}_{s}$ and the untouched long-term cell state $\mathbf{c}^{(k)}_l$ are combined to generate the final decayed cell state $\mathbf{c}^{(k)}_*$ by:
\begin{equation}\label{eq:outdecay2}
    \mathbf{c}^{(k)}_* = \mathbf{c}^{(k)}_l+\hat{\mathbf{c}}^{(k)}_{s},
\end{equation}
which is the adjusted input cell state to the $k$-th LSTM cell, generating the output representation for the $k$-th aggregation layer.

\begin{algorithm}[t]
  \SetKwInOut{Input}{input}\SetKwInOut{Output}{output}
  \Input{Temporal interaction graph edge set $\mathcal{E}$; Aggregation layer $K$; NDP controlling parameter $\lambda$. }
  Initialize all model parameters\; 
  \ForEach{batch $ \subseteq \mathcal{E}$}{
    \ForEach{$k=1,2, ..., K$}{
    Retrieve neighborhood for target node $i$\;
    Compute occurrence-aware attention score by Equation \ref{eq:occurrence-basedscore}\;
    Compute the final score $\hat{\mathbf{a}}_i^{(k)}$ and refine neighborhood information $\widetilde{\mathbf{h}}_i$\;
    Apply outdated-decay by Equation \ref{eq:outdecay1}-\ref{eq:outdecay2}\;
    Round layer state $\hat{u}^{(k)}$ to $u^{(k)}$ as Equation \ref{eq:bernulli}\;
    Aggregate neighborhood by Equation \ref{eq:update}\;
    Update $\hat{u}^{(k+1)}$ by Equation \ref{eq:stateupdate}\;
    }
    Compute NDP loss $\mathcal{L}_{\text{NDP}}$ by Equation \ref{NDPloss}\;
    Compute loss $\mathcal{L}$ with $\lambda$ by Equation \ref{finalloss} and backward\;
  }
  \caption{Traning {\model} (one epoch).}
  \label{algorithm}
\end{algorithm}

\begin{table*}[t]
    \caption{Average Precision (\%) results on Temporal Link Prediction task under both transductive and inductive settings. Results with the \textcolor{pink}{pink background} represent the performance and corresponding improvements achieved by integrating our model into three backbones. The best results are highlighted in bold, and * denotes the benchmark dataset introduced in this paper.} 
    \label{linkprediction}
    \small
    \setlength{\tabcolsep}{4pt}
    \begin{tabular}{cccccc|ccccc}
         \toprule
         & \multicolumn{5}{c}{\textbf{Transductive setting}} & \multicolumn{5}{c}{\textbf{Inductive setting}} \\
            & Wikipedia & Reddit & MOOC & LastFM & TemFin* & Wikipedia & Reddit & MOOC & LastFM & TemFin*\\
            \midrule
             JODIE     & 94.62 $\pm$ 0.5 & 91.11 $\pm$ 0.3 & 76.50 $\pm$ 1.8 & 68.77 $\pm$ 3.0 & 88.02 $\pm$ 0.2 & 93.11 $\pm$ 0.4 & 94.36 $\pm$ 1.1 & 77.83 $\pm$ 2.1 & 82.55 $\pm$ 1.9 & 75.35 $\pm$ 1.6 \\
             TGAT      & 95.34 $\pm$ 0.1 & 98.12 $\pm$ 0.2 & 60.97 $\pm$ 0.3 & 53.36 $\pm$ 0.1 & 87.70 $\pm$ 0.1 & 93.99 $\pm$ 0.3 & 96.62 $\pm$ 0.3 & 63.50 $\pm$ 0.7 & 55.65 $\pm$ 0.2 & 79.37 $\pm$ 0.2 \\
             DyRep     & 94.59 $\pm$ 0.2 & 97.98 $\pm$ 0.1 & 75.37 $\pm$ 1.7 & 68.77 $\pm$ 2.1 & 87.99 $\pm$ 0.1 & 92.05 $\pm$ 0.3 & 95.68 $\pm$ 0.2 & 78.55 $\pm$ 1.1 & 81.33 $\pm$ 2.1 & 76.53 $\pm$ 0.1 \\
             PINT      & 98.78 $\pm$ 0.1 & 99.03 $\pm$ .01 & 85.14 $\pm$ 1.2 & 88.06 $\pm$ 0.7 & 90.79 $\pm$ 0.1 & 98.38 $\pm$ .04 & 98.25 $\pm$ .04 & 85.39 $\pm$ 1.0 & 91.76 $\pm$ 0.7 & 81.18 $\pm$ 0.2 \\
             iLoRE     & 98.98 $\pm$ 0.3 & 99.11 $\pm$ 0.4 & 90.44 $\pm$ 1.0 & 91.39 $\pm$ 0.1 & 90.90 $\pm$ 0.1 & 98.60 $\pm$ 0.3 & 98.65 $\pm$ 0.3 & 89.75 $\pm$ 0.8 & 93.29 $\pm$ 0.8 & 84.33 $\pm$ 0.2 \\
             GraphMixer& 97.95 $\pm$ .03 & 97.31 $\pm$ .01 & 82.78 $\pm$ 0.2 & 67.27 $\pm$ 2.1 & 86.85 $\pm$ 0.1 & 96.65 $\pm$ .02 & 95.26 $\pm$ .02 & 81.41 $\pm$ 0.2 & 82.11 $\pm$ 0.4 & 77.47 $\pm$ 0.2 \\
             \midrule
             TGN       & 98.46 $\pm$ 0.1 & 98.70 $\pm$ 0.1 & 85.88 $\pm$ 3.0 & 80.69 $\pm$ 0.2 & 90.65 $\pm$ 0.1 & 97.81 $\pm$ 0.1 & 97.55 $\pm$ 0.1 & 85.55 $\pm$ 2.9 & 84.66 $\pm$ 0.1 & 80.67 $\pm$ 0.2 \\
             \rowcolor{myPink}\textbf{TGN+}     & 98.79 $\pm$ 0.1 & 98.73 $\pm$ .02 & 89.91 $\pm$ 1.2 & 84.15 $\pm$ 2.5 & 91.04 $\pm$ 0.1 & 98.18 $\pm$ 0.1 & 97.69 $\pm$ 0.1 & 88.91 $\pm$ 1.6 & 88.60 $\pm$ 1.2 & 81.30 $\pm$ 0.6 \\
             \rowcolor{myPink}(\textbf{\textit{imprv.}}) & (\textit{+0.33}) & (\textit{+0.03}) & (\textit{+4.03}) & (\textit{+3.46}) & (\textit{+0.39}) & (\textit{+0.37}) & (\textit{+0.14}) & (\textit{+3.36}) & (\textit{+3.94}) & (\textit{+0.63})\\
             \midrule
             TIGE      & 98.38 $\pm$ 0.1 & 99.04 $\pm$ 0.1 & 89.64 $\pm$ 0.9 & 87.85 $\pm$ 0.9 & 90.81 $\pm$ .02 & 98.45 $\pm$ 0.1 & 98.39 $\pm$ 0.1 & 89.51 $\pm$ 0.7 & 90.14 $\pm$ 1.0 & 83.22 $\pm$ 0.2 \\
             \rowcolor{myPink}\textbf{TIGE+}     & 98.95 $\pm$ .04 & 99.13 $\pm$ .02 & \textbf{92.26 $\pm$ 0.8} & 91.42 $\pm$ 0.7 & 90.92 $\pm$ .01 & 98.61 $\pm$ 0.1 & 98.50 $\pm$ .03 & \textbf{91.53 $\pm$ 0.8} & 93.14 $\pm$ 0.6 & 83.79 $\pm$ 0.1 \\
             \rowcolor{myPink}(\textbf{\textit{imprv.}}) & (\textit{+0.12}) & (\textit{+0.09}) & \textbf{(\textit{+2.62})} & (\textit{+3.79}) & (\textit{+0.11}) & (\textit{+0.16}) & (\textit{+0.11}) & \textbf{(\textit{+2.02})} & (\textit{+3.08}) & (\textit{+0.57}) \\
             \midrule
             DyGFormer  & 99.03 $\pm$ .02 & 99.22 $\pm$ .01 & 87.52 $\pm$ 0.5 & 93.00 $\pm$ 0.1 & 93.34 $\pm$ 0.2 & 98.59 $\pm$ .03 & 98.84 $\pm$ .02 & 86.96 $\pm$ 0.4 & 94.23 $\pm$ .09 & 86.29 $\pm$ 0.4 \\
             \rowcolor{myPink}\textbf{DyGFormer+} & \textbf{99.27 $\pm$ .02} & \textbf{99.59 $\pm$ .01} & 88.29 $\pm$ 0.5 & \textbf{94.73 $\pm$ 0.1} & \textbf{95.81 $\pm$ 0.3} & \textbf{98.63 $\pm$ .04} & \textbf{98.91 $\pm$ 0.1} & 87.20 $\pm$ 0.5 & \textbf{94.89 $\pm$ 0.1} & \textbf{86.91 $\pm$ .04} \\
             \rowcolor{myPink}(\textbf{\textit{imprv.}}) & \textbf{(\textit{+0.24})} & \textbf{(\textit{+0.37})} & (\textit{+0.77}) & \textbf{(\textit{+1.73})} & \textbf{(\textit{+2.47})} & \textbf{(\textit{+0.04})} & \textbf{(\textit{+0.07})} & (\textit{+0.24}) & \textbf{(\textit{+0.66})} & \textbf{(\textit{+0.62})} \\
            \bottomrule
    \end{tabular}
    \label{tab:my_label}
\end{table*}
\begin{table*}[t]
  \caption{AUROC (\%) results on Evolving Node Classification task. Due to table restrictions, we omit some baseline results that have been shown inferior to our three backbones \cite{tgn,tiger,dygformer}.}
  \small
  \setlength{\tabcolsep}{2pt}
  \label{evolvingClassification}
  \begin{tabular}{cccc|c>{\columncolor{myPink}}c>{\columncolor{myPink}}c|c>{\columncolor{myPink}}c>{\columncolor{myPink}}c|c>{\columncolor{myPink}}c>{\columncolor{myPink}}c}
    \toprule
      & DyRep & PINT & GraghMixer & TGN & \textbf{TGN+} & (\textbf{\textit{imprv.}}) & TIGE & \textbf{TIGE+} & (\textbf{\textit{imprv.}}) & DyGFormer & \textbf{DyGFormer+} & (\textbf{\textit{imprv.}})   \\
    \midrule
    Wikipedia & 84.59 $\pm$ 2.2 & 87.59 $\pm$ 0.6 & 86.80 $\pm$ .01 & 87.81 $\pm$ 0.3 & \textbf{87.91 $\pm$ 0.8} & \textbf{(\textit{+0.10})} & 86.92 $\pm$ 0.7 & 87.16 $\pm$ 1.5 & (\textit{+0.24}) & 87.07 $\pm$ 0.8 & 87.49 $\pm$ 0.6 & (\textit{+0.42}) \\
    Reddit    & 62.91 $\pm$ 2.4 & 67.31 $\pm$ 0.2 & 64.22 $\pm$ .03 & 67.06 $\pm$ 0.9 & 68.26 $\pm$ 0.8 & (\textit{+1.20}) & 69.41 $\pm$ 1.3 & \textbf{70.24 $\pm$ 2.2} & \textbf{(\textit{+0.83})} & 68.30 $\pm$ 1.5 & 69.06 $\pm$ 2.5 & (\textit{+0.76})  \\
    MOOC      & 67.76 $\pm$ 0.5 & 68.77 $\pm$ 1.1 & 67.21 $\pm$ .02 & 69.54 $\pm$ 1.0 & 72.96 $\pm$ 1.8 & (\textit{+2.04}) & 72.35 $\pm$ 2.3 & 73.85 $\pm$ 2.0 & (\textit{+1.50}) & 77.89 $\pm$ 0.5 & \textbf{78.88 $\pm$ 3.2} & \textbf{(\textit{+0.99})}   \\
    TemFin*   & 78.56 $\pm$ 2.9 & 82.22 $\pm$ 1.2 & 81.17 $\pm$ 0.7 & 80.93 $\pm$ 2.2 & 82.39 $\pm$ 2.5 & (\textit{+1.46}) & 80.52 $\pm$ 3.0 & \textbf{83.17 $\pm$ 2.2} & \textbf{(\textit{+2.65})} & 72.50 $\pm$ 0.4 & 73.05 $\pm$ 2.1 & (\textit{+0.55}) \\
    \bottomrule
  \end{tabular}
\end{table*}

\subsection{Training}
\subsubsection{Error Gradients}
The entire model is differentiable except for the round process $f_\text{round}(\cdot)$ in Equation \ref{eq:bernulli}. In this paper, we employ the straight-through estimator \cite{skiprnn, dicretevaribale1} to allow all parameters to be trained efficiently without the need for any extra supervision signals. This involves approximating the step process with identity during gradient computation in the backward pass: $\frac{\partial f_{\text {round}}(x)}{\partial x}=1$. 

\subsubsection{Loss Function}
We employ temporal link prediction as our self-supervised task for training. Specifically, for each link $(i, j, t)$, we calculate its occurrence probability $\hat{p}_{ij}(t)$ with the concatenated temporal representations of interaction nodes, \ie, $\mathbf{z}_i(t)$ and $\mathbf{z}_j(t)$, through a two-layer MLP \cite{tiger}. Then, we compute the loss function by the cross entropy of the link prediction task as follows:
\begin{equation}\label{eq:negative}
    \mathcal{L}_\text{link}=-\sum_{(i, j, t) \in \mathcal{E}}\left[\log \hat{p}_{i j}(t)+\log \left(1-\hat{p}_{i k}(t)\right)\right],
\end{equation}
where $k$ is the negative destination node by random sampling. Finally, we consider our NDP loss $\mathcal{L}_\text{NDP}$ in Equation \ref{NDPloss} for regularizing the learned attention mechanisms in our final training loss:
\begin{equation}\label{finalloss}
    \mathcal{L} = \mathcal{L}_\text{link} + \lambda \mathcal{L}_\text{NDP},
\end{equation}
where $\lambda \in \mathbb{R}^+$ is a controlling hyper-parameter for our penalty method.

\subsection{Complexity Analysis}
We provide a detailed complexity analysis of SEAN where we show the complexity of both components. The Representative Neighbor Selector module assigns distinct attention scores to nodes and also provides the neighbor diversity penalty loss, leading to the computational complexity of $\mathcal{O}(K|\mathcal{V}|d)$, where $K, |\mathcal{V}|, d$ refer to the total number of stacking layers, nodes, and the embedding dimension, respectively. The Temporal-aware Aggregator is implemented based on the LSTM module, therefore leading to a complexity of $\mathcal{O}(K|\mathcal{V}|d^2)$. Therefore, the overall complexity of SEAN is $\mathcal{O}(K|\mathcal{V}|d + K|\mathcal{V}|d^2)$, maintaining efficiency as it scales \textit{linearly} with the number of nodes.

We want to emphasize that the temporal embedding module mentioned in Section \ref{sec:temporalembeddingmodule} in most existing TGNs also incorporates an explicit attention assignment stage. In this case, integrating SEAN \textit{only} leads to an additional overall complexity of $\mathcal{O}(K|\mathcal{V}|d^2)$.

\section{Experiments}
\subsection{Experimental Settings} \label{sec:datasets}
\subsubsection{Datasets.}
We conduct experiments with five datasets, including four public datasets \cite{jodie} and one dataset introduced in our paper. The four public datasets - Wikipedia, Reddit, MOOC, and LastFM - are widely used in temporal interaction graph modeling. 
Meanwhile, recognizing that these available datasets are social or event networks, we release TemFin, a new TIG benchmark dataset from the complicated financial domain. TemFin consists of half a month of transactions sampled from a private financial transfer transaction network in the Ant Finance Group.%
\footnote{The dataset is sampled properly only for experiment purposes and does not imply any commercial information. All personal identity information (PII) has been removed.} Details of all datasets are described in Table \ref{datasets} in the Appendix due to the page limitations. All datasets are sequentially split according to the edge timestamp order by $70\%$, $15\%$, and $15\%$ for training, validation, and testing, respectively~\cite{tiger}. 

\subsubsection{Baselines.}
For comparison, we choose nine existing TGNs as our baselines, and summarize the detailed description of our baselines as follows:
\begin{itemize}[leftmargin=*]
\item \textbf{DyRep} \cite{dyrep}: It is a notable implementation of the temporal point process in its neighborhoods and introduces a projection layer that estimates user representation in the future.
\item \textbf{JODIE} \cite{jodie}: This method considers the dynamic representation of nodes changing over time and utilizes two RNNs for temporal interaction graph modeling.
\item \textbf{TGAT} \cite{tgat}: TGAT aggregates its neighborhoods by applying a temporal embedding module based on the temporal attention mechanism.
\item \textbf{TGN} \cite{tgn}: TGN summarizes the former models and proposes a memory module to record the historical behaviors of nodes, significantly improving the performance of temporal interaction graph modeling.
\item \textbf{TIGE} \cite{tiger}: TIGE enhances the TGN framework by incorporating a dual memory module, effectively addressing the issue of staleness and building upon TGN's foundational concepts.
\item \textbf{GraphMixer} \cite{graphmixer}: This approach employs a simple MLP-Mixer for neighborhood encoding and has demonstrated impressive performance in the Temporal Link Prediction task under the transductive setting.
\item \textbf{PINT} \cite{pint}: PINT utilizes injective temporal message passing on neighborhoods and leverages relative positional features to improve model performance.
\item \textbf{iLoRE} \cite{ilore}: Focusing on instant long-term modeling and re-occurrence preservation, iLoRE provides a nuanced approach to TIG modeling.
\item \textbf{DyGFormer} \cite{dygformer}: This model introduces a novel architecture that performs Transformer blocks through the patch technique, achieving SOTA results in TIG modeling.
\end{itemize}


We select three representative baselines to serve as the backbones for our {\model}: TGN \cite{tgn}, TIGE \cite{tiger}, and DyGFormer \cite{dygformer}. We then integrate our model into these three backbones and evaluate their performance in two downstream tasks, \ie, Temporal Link Prediction and Evolving Node Classification.

\begin{figure*}
    \centering
    \includegraphics[width=\linewidth]{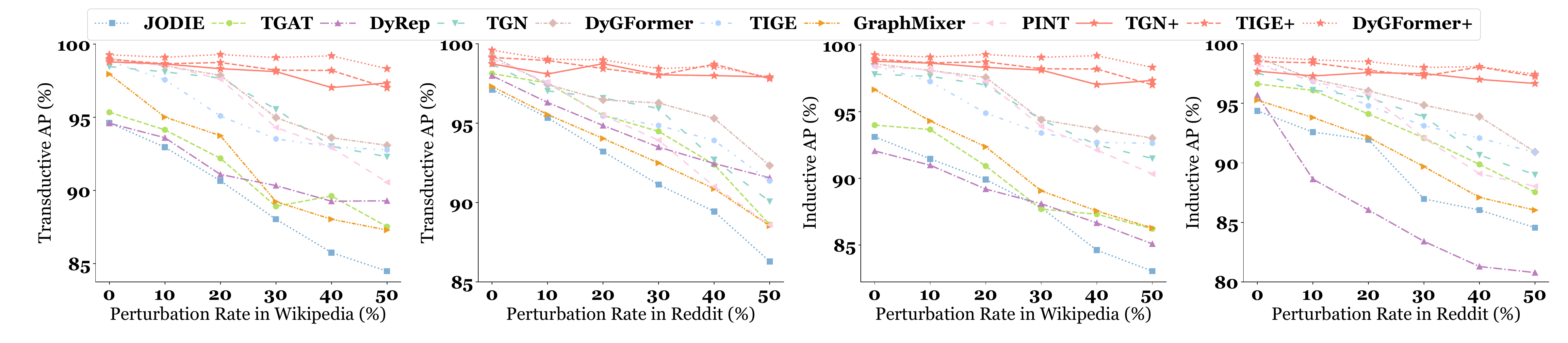}
    \caption{Robustness to noisy neighborhoods on Wikipedia and Reddit in different perturbation rates.}
    \label{fig:noise_wiki_reddit}
\end{figure*}

\begin{figure}
    \includegraphics[width=\linewidth]{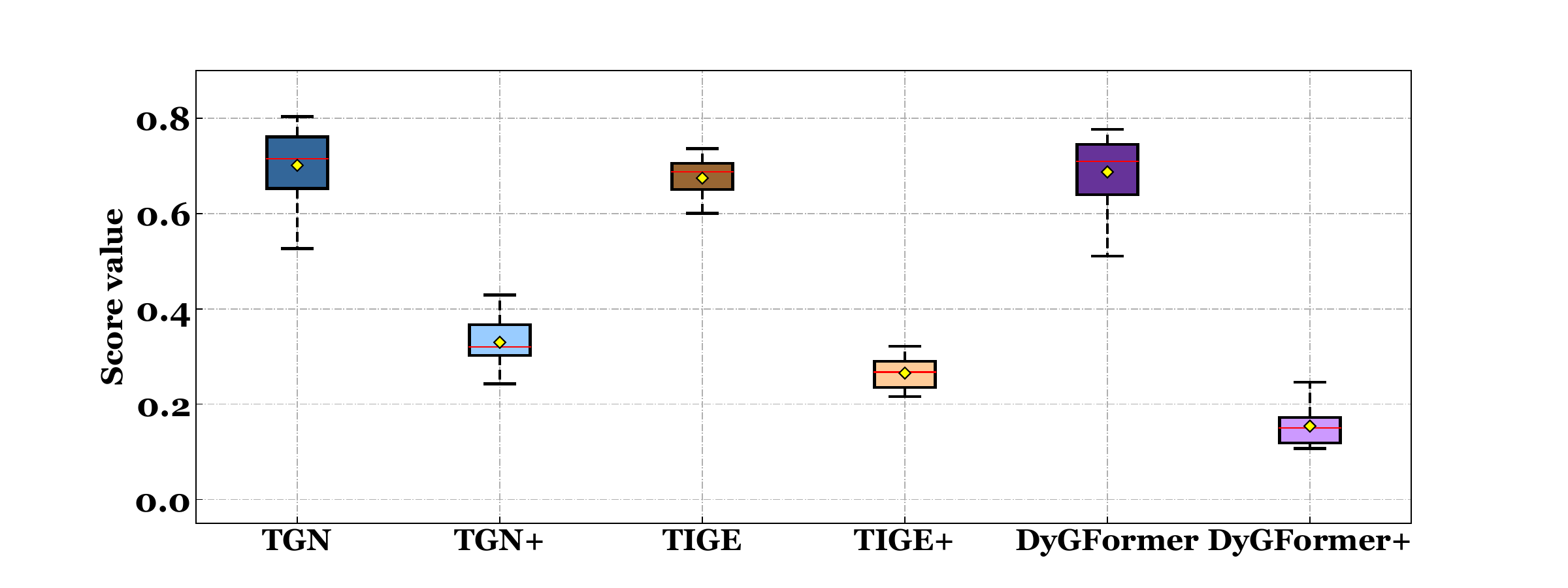}
    \caption{Attention scores assigned from randomly selected nodes to their perturbed neighbors on noisy Wikipedia.}
    \label{fig:noise_wiki_score}
\end{figure}

\subsection{Temporal Link Prediction}\label{sec:link}
We start our experiments with the Temporal Link Prediction task, which aims to predict the probability of a link occurring between two specific nodes at a certain time. We utilize two settings: the transductive setting, which performs link prediction on nodes that have appeared during training, and the inductive setting, which predicts links between unseen nodes. We randomly sample an equal number of negative nodes as detailed in Equation \ref{eq:negative}, and report the average precision (AP) performance with all three backbones, \ie, TGN \cite{tgn}, TIGE \cite{tiger}, and DyGFormer \cite{dygformer}. 

The results are shown in Table \ref{linkprediction}. Clearly, we can observe that all three backbone models show improved performance across all datasets in both transductive and inductive settings after integrating our model. Meanwhile, among the improved models, at least one has achieved SOTA results, surpassing all baseline comparisons. This observation proves the effectiveness of the adaptive neighborhood encoding mechanism in TGNs. Moreover, we find the most significant improvement between the backbones and their corresponding improved models specifically on MOOC and lastFM. This may be owing to their high neighborhood complexity, constraining the backbones' effectiveness. In contrast, our model introduces an adaptive neighborhood encoding mechanism that effectively empowers these backbones to better capture and utilize the critical information in such challenging environments, demonstrating our model's effectiveness and robustness in handling complex TIGs.

\subsection{Evolving Node Classification} \label{sec:node}
We also conduct the Evolving Node Classification task as a downstream task to further validate the effectiveness of our learned temporal node representations. Specifically, we input the learned representation of node $i$ at time $t$, $\mathbf{z}_i(t)$, into a two-layer MLP, which maps the representations to the dynamic labels. We carry out the experiments on Wikipedia, Reddit, MOOC, and TemFin. LastFM is excluded due to its lack of node labels. We employ AUROC as our evaluation metric due to the label imbalance issue in these datasets.

We report the results in Table \ref{evolvingClassification}. All three improved models demonstrate better performance than the backbones across all datasets, and at least one of them has achieved the SOTA results. This indicates that the learned representations from our model can be more effectively applied to downstream tasks, proving its superiority once again. Notably, the most significant improvement is observed on MOOC and TemFin. This could be due to their relatively less severe label imbalance issue compared to others, thereby facilitating the performance enhancement more readily.

\begin{figure}
  \centering
  \begin{subfigure}{.49\linewidth}
    \centering
    \includegraphics[width=0.49\linewidth]{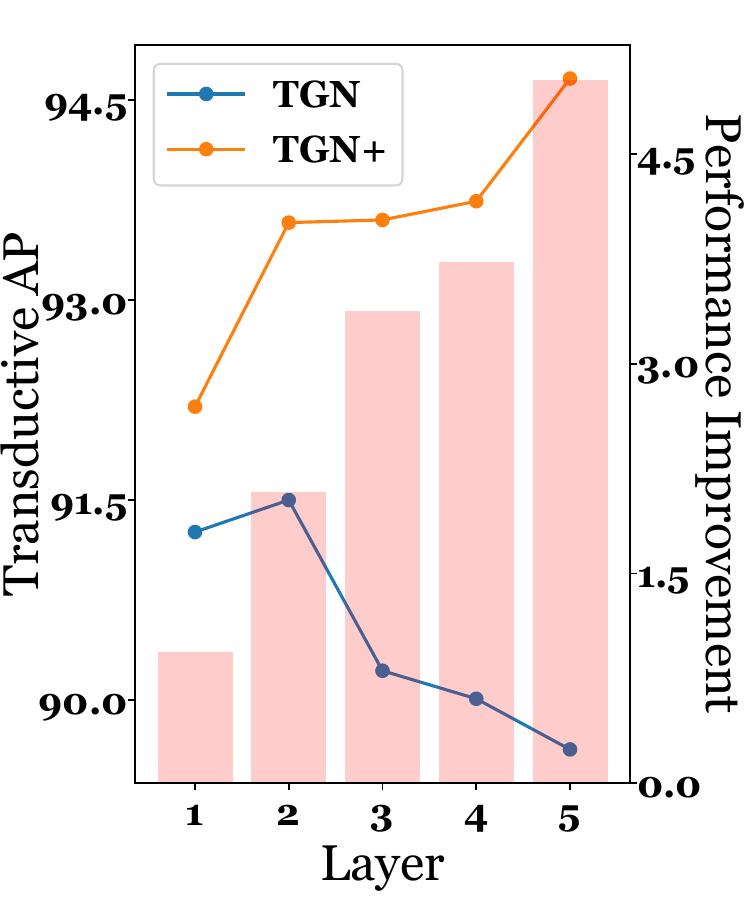}
    \includegraphics[width=0.49\linewidth]{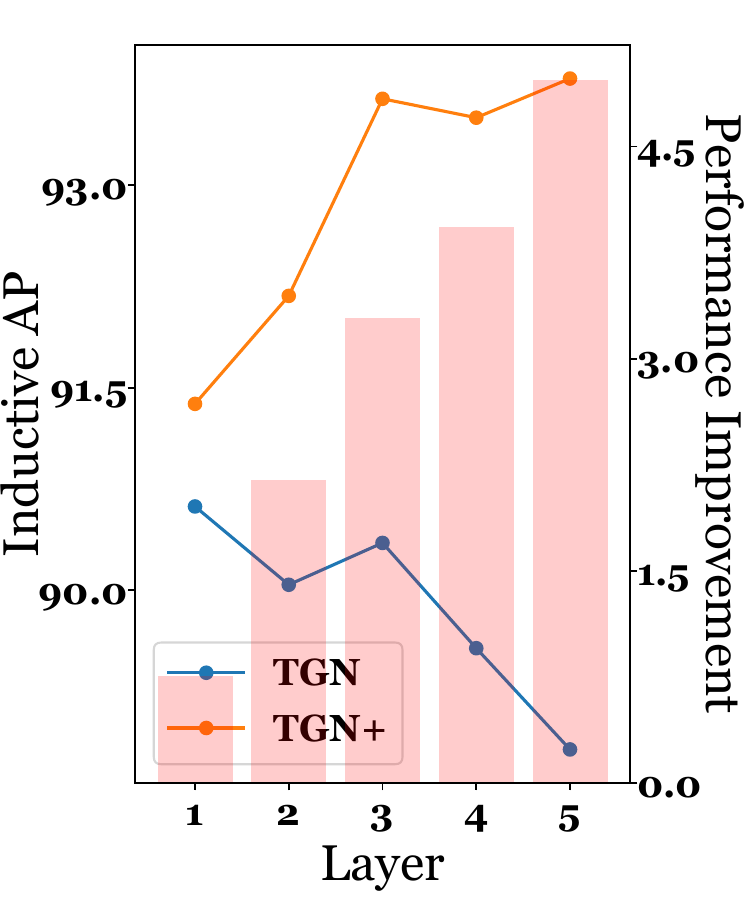}
    \caption{TGN model.}
  \end{subfigure}
  \centering
  \begin{subfigure}{.49\linewidth}
    \centering
    \includegraphics[width=.49\linewidth]{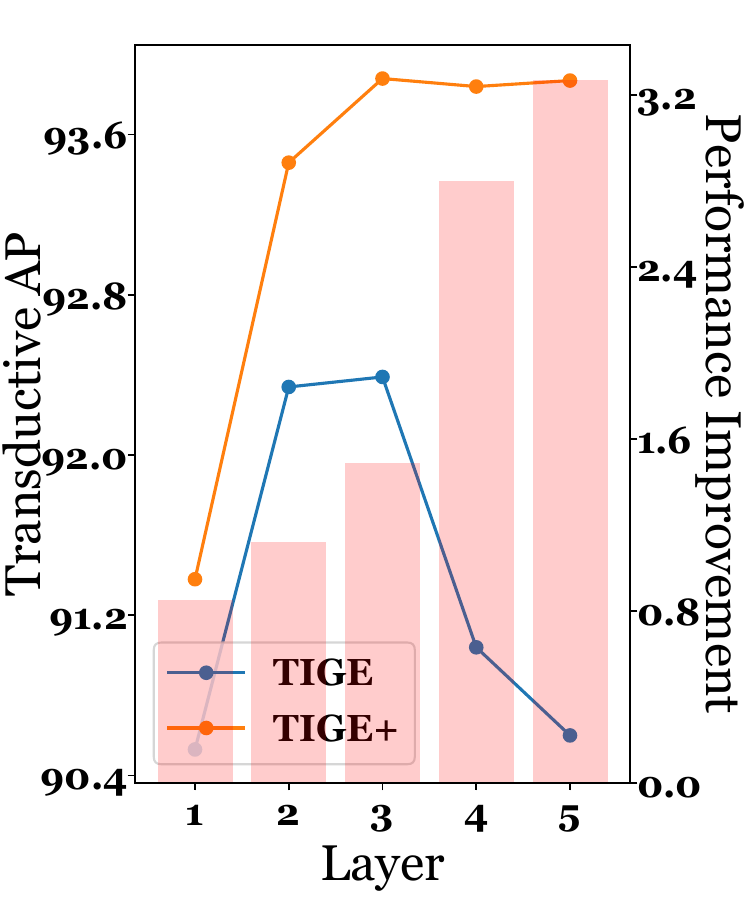}
    \includegraphics[width=.49\linewidth]{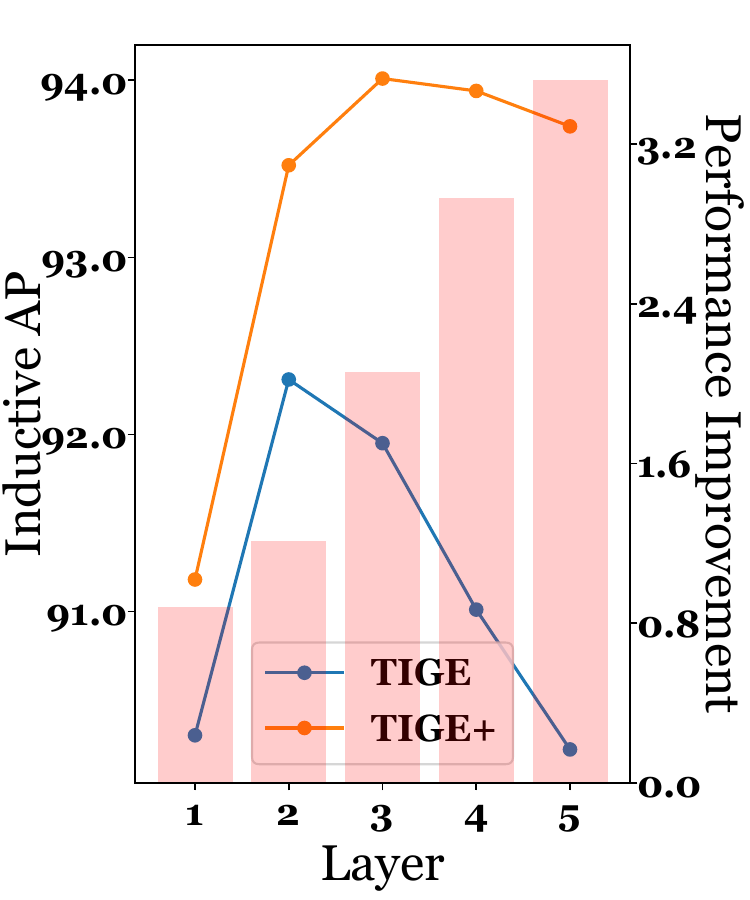}
    \caption{TIGE model.}
  \end{subfigure}
  \caption{Robustness to expanded neighborhoods on MOOC.}
  \label{fig:layer}
\end{figure}

\subsection{Robustness to Noisy Neighborhoods} \label{sec:noisydata}
We have mentioned that the suitable neighborhoods for nodes should vary due to the different connectivity patterns and potential noise. To simulate this scenario, we complicate and pollute the nodes' neighborhoods by introducing random noise into TIGs \cite{rdgsl}. We then evaluate the model performance (both w/o and w/ {\model}) under these noisy conditions. Specifically, we replace the original links with our randomly sampled noisy links at a certain perturbation rate $p \in \{10\%, 20\%, 30\%, 40\%, 50\%\}$. Meanwhile, at $p=50\%$, we collect the attention scores assigned from a subset of randomly selected nodes to their perturbed neighbors, visualizing the model's ability to handle noise with box plots. We emphasize that these comparisons are controlled and fair because the noise issue dose exist in TIGs and the introduced noise is prevalent \cite{rdgsl, mata}.

The results are detailed in Figure \ref{fig:noise_wiki_reddit} and \ref{fig:noise_wiki_score}. The improved models exhibit exceptional robustness, even when encountering significant noise. This demonstrates the superiority of {\model} in encoding suitable neighborhoods, thus weakening potential noise attacks. As for the perturbed neighbors' scores, we find an obvious reduction and greater focus in the scores when comparing improved models to their corresponding backbones. It suggests that {\model} successfully empowers these backbones to evaluate and mitigate these same-level random noises introduced in nodes' neighborhoods.

\subsection{Robustness to Expanded Neighborhoods} \label{sec:layer}
We compare the performance across expanded neighborhoods between the backbones and improved models. Specifically, we vary the aggregation layer $K \in \{1, 2, 3, 4, 5\}$ in models and carry out the experiments using TGN \cite{tgn} and TIGE \cite{tiger} as the backbones. DyGFormer \cite{dygformer} is not included in this group of experiments due to its limitation to the first-hop neighborhoods in the original design. In Section \ref{linkprediction}, we keep the neighbor sampling size of TGN \cite{tgn} at 10 to ensure a fair comparison, which is its default hyper-parameter. However, during these experiments, maintaining the sampling size at 10 will lead to GPU out-of-memory issues rapidly. Therefore, we reduce the size to 5 in practice.

The results are depicted in Figure \ref{fig:layer}. The improved models also exhibit significant robustness with expanded neighborhoods, and the performance improvement compared to their corresponding backbones progressively widens as the neighborhood expands. It highlights {\model}'s capability to adaptively encode the suitable neighborhoods, thus avoiding the over-smoothing issue \cite{over-smoothing} caused by the expanded neighborhoods. Moreover, fine-tuning the accessible neighborhoods (\ie, the layer number $K$) is essential for optimizing the backbone performance. In contrast, our improved models reach their peak performance by employing the largest possible accessible neighborhoods, as long as the computational resources permit.

\subsection{Case Study}\label{sec:case}
\begin{figure}
    \centering
    \includegraphics[width=\linewidth]{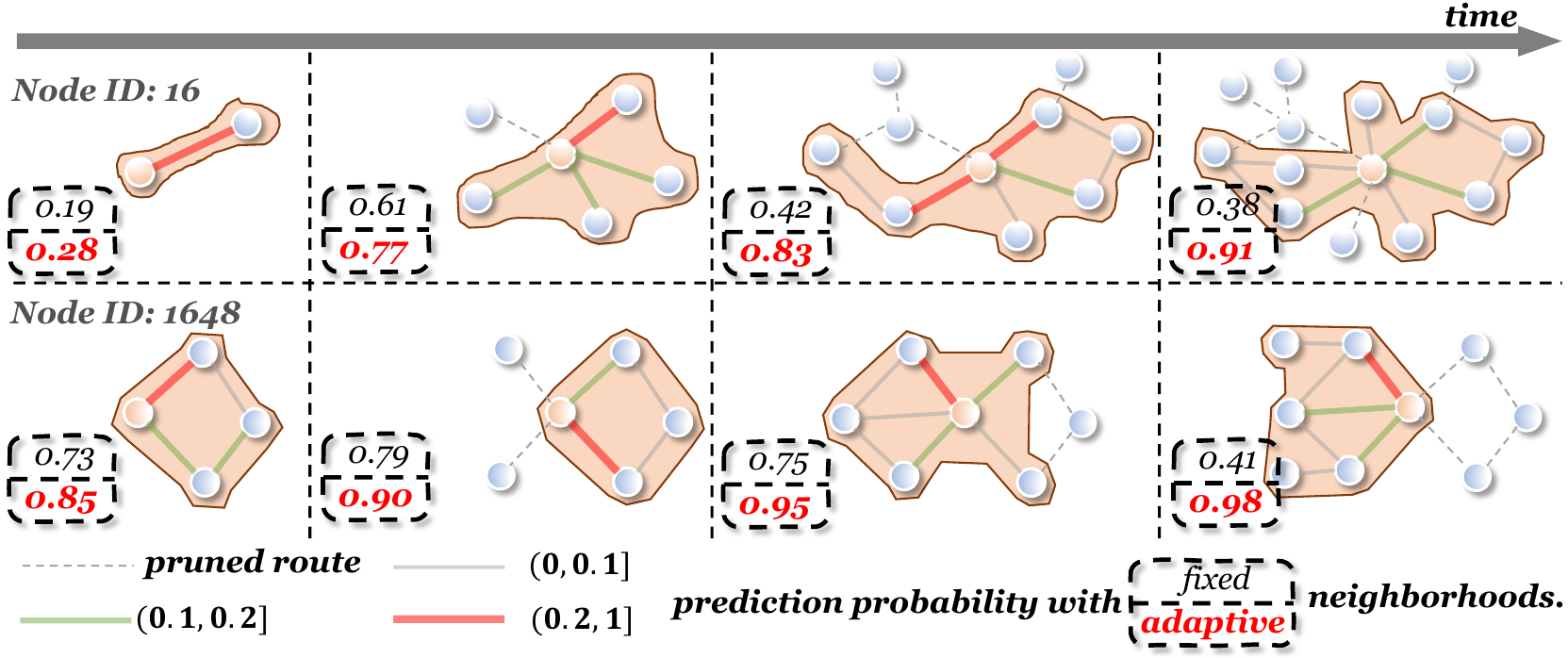}
  \caption{Case study on Wikipedia with TGN model. The introduction of {\model} can lead to both adaptive neighborhood consideration and improved prediction performance.}
  \label{fig:casestudy}
\end{figure}

We conduct the case study to interpret the choice-making process and model effectiveness when {\model} seeks the neighborhoods. As shown in Figure \ref{fig:casestudy}, each grid displays the encoded neighborhood for a certain node at a particular time, and dashed lines indicate the aggregation routes pruned by {\model}. The values within those dashed boxes are the Evolving Node Classification prediction probabilities for the backbone and the improved model, respectively. Here, higher values signify better performance. For clarity, we limit the depiction only to 2-hop accessible neighborhoods for each node.

In our observations from Node 16 in Figure \ref{fig:casestudy}, the backbone does not consistently show improved performance over time. Conversely, the improved model performance continues to increase, indicating {\model}'s ability to manage the increasing neighborhood complexity and progressively stronger potential noise. Another notable aspect is that, even within the same node such as Node 1648 in Figure \ref{fig:casestudy}, the improved model shows significant neighborhood differences at distinct timestamps. It suggests that the improved model can effectively encode the most suitable neighborhoods across different timestamps to achieve optimal performance.

\subsection{Ablation Study}\label{sec:ablation}
We conduct the ablation study on the main components of our {\model}, including Representative Neighbor Selector (RNS) in Section \ref{sec:bs}, Temporal-aware Aggregator (TA) in Section \ref{sec:DS}, Neighbor Diversity Penalty (NDP) in Section \ref{sec:NDP}, Adaptive Pruning Module (APM) in Section \ref{sec:prune}, and Outdated-decay Module (OM) in Section \ref{timedecay}. To assess the impact of each component, we remove them individually, resulting in five variants: w/o RNS, w/o TA, w/o NDP, w/o APM, and w/o OM.

\begin{figure}
    \centering
    \includegraphics[width=\linewidth]{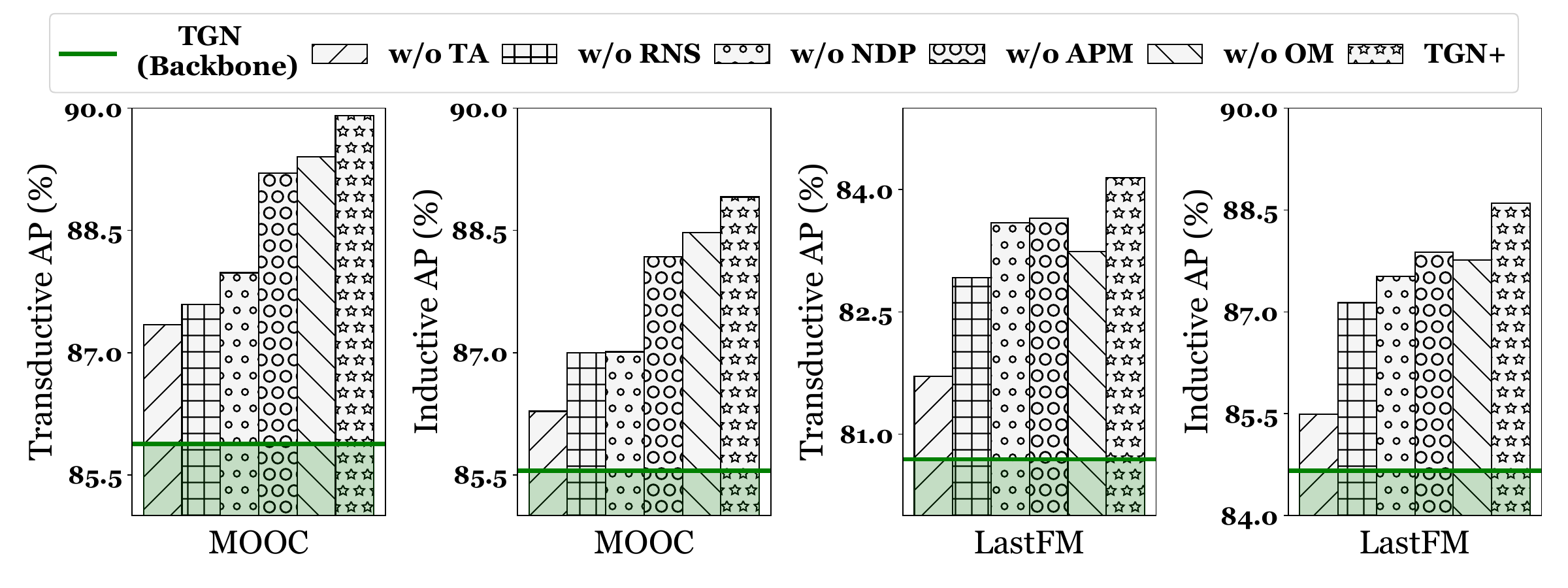}
    \caption{Ablation study on MOOC and LastFM with the improved TGN model. TGN backbone is shown for reference.}
    \label{fig:ablation}
\end{figure}

We report the performance of the Temporal Link Prediction task using both the improved TGN model and its variants as represented in Figure \ref{fig:ablation}. Note that the green line represents the performance of the TGN backbone. The model performs best when utilizing all components, and the removal of any single component leads to a decrease in performance, highlighting the necessity of each design. Moreover, the performance of all variants surpasses the backbone, validating the effectiveness of our components. Additionally, our Temporal-aware Aggregator component has the most significant contribution to model performance, reaffirming the importance of the temporal-aware design for neighborhood encoding.

\subsection{Parameter Study}\label{sec:parameter}
\begin{figure}
  \centering
  \begin{subfigure}{.49\linewidth}
    \centering
    \includegraphics[width=\linewidth]{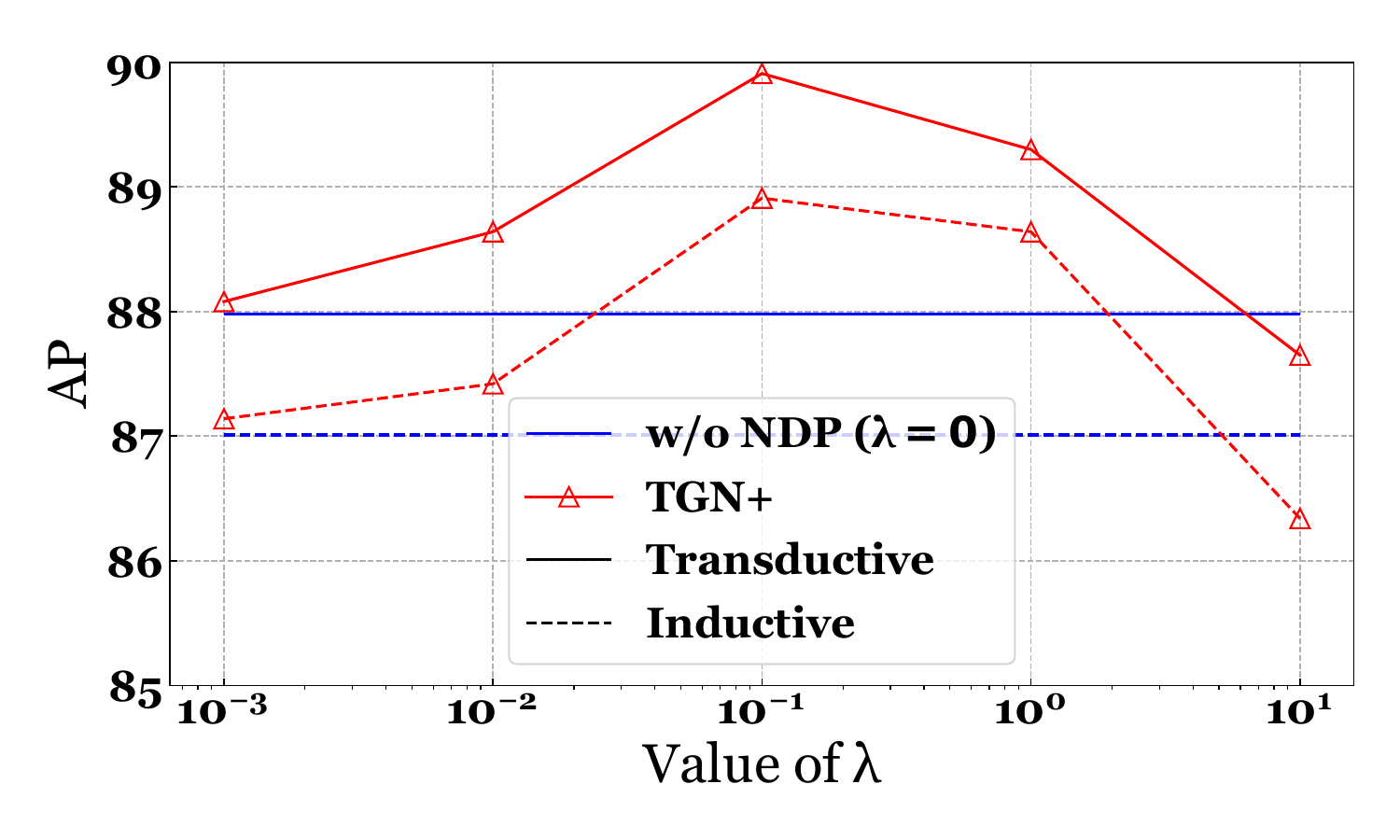}
  \end{subfigure}
  \centering
  \begin{subfigure}{.49\linewidth}
    \centering
    \includegraphics[width=\linewidth]{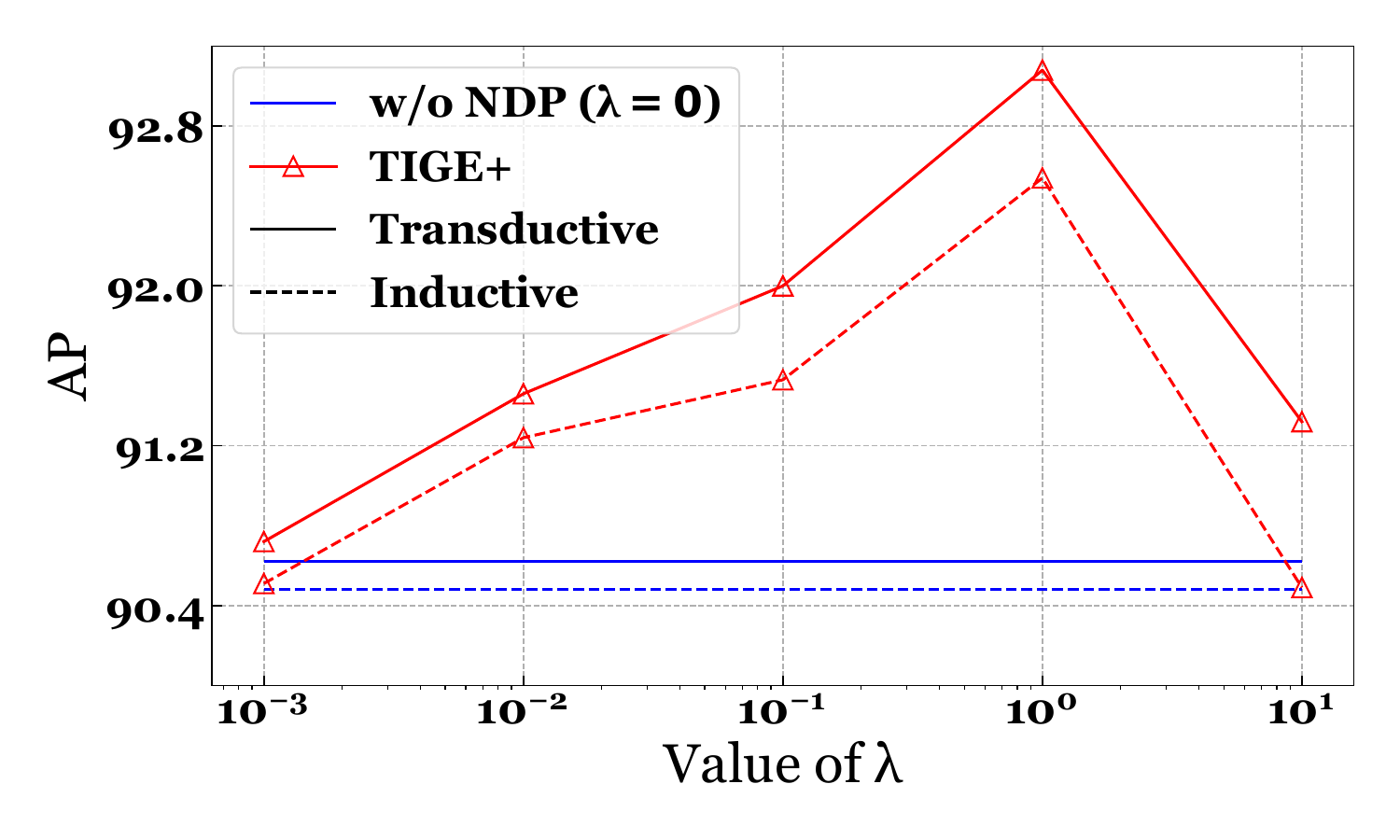}
  \end{subfigure}
  \caption{Parameter study on MOOC with improved TGN (left) and TIGE (right) models. w/o NDP is shown for~reference.}
  \label{fig:para}
\end{figure}

Now, we study how the controlling hyper-parameter of the Neighbor Diversity Penalty ($\lambda$ in Equation \ref{finalloss}) affects the performance. We plot the APs of the improved models with varying values of $\lambda$ on MOOC in Figure \ref{fig:para}. When the penalty method is too strong ($\lambda=10$), the model may struggle to learn information from those important neighbors, thus leading to a decrease in model performance. Additionally, different models have distinct sensitivities to $\lambda$ changes. For example, TIGE requires a larger $\lambda$ for optimal results compared to TGN, suggesting a more severe over-concentration issue within the TIGE model.

\section{Conclusion and future work}
In this paper, we enhance temporal interaction graph modeling via adaptive neighborhood encoding. We propose {\model}, a plug-and-play model designed to boost existing TGNs. By introducing the Representative Neighbor Selector, we can select the personalized representatives for each target node. The Temporal-aware Aggregator then aggregates neighborhoods in a temporal-aware manner. As for future work, we will consider detecting the hidden routes that do not originally exist during the aggregation process. Additionally, we can also consider other implements for our Temporal-aware Aggregator, such as Transformer blocks.

\begin{acks}

This work is funded in part by the Shanghai Science and Technology Development Fund No.22dz1200704, the National Natural Science Foundation of China Projects No. U1936213, NSF through grants IIS-1763365 and IIS-2106972, and also supported by Ant Group. The first author, Dr. Zhang, also wants to thank Yifeng Wang for his efforts and support in this work.
\end{acks}



\clearpage
\newpage

\bibliographystyle{ACM-Reference-Format}
\bibliography{sample-base}


\begin{thebibliography}{47}


\ifx \showCODEN    \undefined \def \showCODEN     #1{\unskip}     \fi
\ifx \showDOI      \undefined \def \showDOI       #1{#1}\fi
\ifx \showISBNx    \undefined \def \showISBNx     #1{\unskip}     \fi
\ifx \showISBNxiii \undefined \def \showISBNxiii  #1{\unskip}     \fi
\ifx \showISSN     \undefined \def \showISSN      #1{\unskip}     \fi
\ifx \showLCCN     \undefined \def \showLCCN      #1{\unskip}     \fi
\ifx \shownote     \undefined \def \shownote      #1{#1}          \fi
\ifx \showarticletitle \undefined \def \showarticletitle #1{#1}   \fi
\ifx \showURL      \undefined \def \showURL       {\relax}        \fi
\providecommand\bibfield[2]{#2}
\providecommand\bibinfo[2]{#2}
\providecommand\natexlab[1]{#1}
\providecommand\showeprint[2][]{arXiv:#2}

\bibitem[\protect\citeauthoryear{Bengio, L{\'e}onard, and Courville}{Bengio et~al\mbox{.}}{2013}]%
        {dicretevaribale1}
\bibfield{author}{\bibinfo{person}{Yoshua Bengio}, \bibinfo{person}{Nicholas L{\'e}onard}, {and} \bibinfo{person}{Aaron Courville}.} \bibinfo{year}{2013}\natexlab{}.
\newblock \showarticletitle{Estimating or propagating gradients through stochastic neurons for conditional computation}.
\newblock \bibinfo{journal}{\emph{arXiv preprint arXiv:1308.3432}} (\bibinfo{year}{2013}).
\newblock


\bibitem[\protect\citeauthoryear{Campos, Jou, Gir{\'o}-i Nieto, Torres, and Chang}{Campos et~al\mbox{.}}{2017}]%
        {skiprnn}
\bibfield{author}{\bibinfo{person}{V{\'\i}ctor Campos}, \bibinfo{person}{Brendan Jou}, \bibinfo{person}{Xavier Gir{\'o}-i Nieto}, \bibinfo{person}{Jordi Torres}, {and} \bibinfo{person}{Shih-Fu Chang}.} \bibinfo{year}{2017}\natexlab{}.
\newblock \showarticletitle{Skip rnn: Learning to skip state updates in recurrent neural networks}.
\newblock \bibinfo{journal}{\emph{arXiv preprint arXiv:1708.06834}} (\bibinfo{year}{2017}).
\newblock


\bibitem[\protect\citeauthoryear{Chen, Liao, Xiong, Zhang, Zhang, Zhang, and Sun}{Chen et~al\mbox{.}}{2023}]%
        {speed}
\bibfield{author}{\bibinfo{person}{Xi Chen}, \bibinfo{person}{Yongxiang Liao}, \bibinfo{person}{Yun Xiong}, \bibinfo{person}{Yao Zhang}, \bibinfo{person}{Siwei Zhang}, \bibinfo{person}{Jiawei Zhang}, {and} \bibinfo{person}{Yiheng Sun}.} \bibinfo{year}{2023}\natexlab{}.
\newblock \showarticletitle{SPEED: Streaming Partition and Parallel Acceleration for Temporal Interaction Graph Embedding}.
\newblock \bibinfo{journal}{\emph{arXiv preprint arXiv:2308.14129}} (\bibinfo{year}{2023}).
\newblock


\bibitem[\protect\citeauthoryear{Chen, Zhang, Xiong, Wu, Zhang, Sun, Zhang, Zhao, and Kang}{Chen et~al\mbox{.}}{2024}]%
        {promptTIG}
\bibfield{author}{\bibinfo{person}{Xi Chen}, \bibinfo{person}{Siwei Zhang}, \bibinfo{person}{Yun Xiong}, \bibinfo{person}{Xixi Wu}, \bibinfo{person}{Jiawei Zhang}, \bibinfo{person}{Xiangguo Sun}, \bibinfo{person}{Yao Zhang}, \bibinfo{person}{Yinglong Zhao}, {and} \bibinfo{person}{Yulin Kang}.} \bibinfo{year}{2024}\natexlab{}.
\newblock \showarticletitle{Prompt Learning on Temporal Interaction Graphs}.
\newblock \bibinfo{journal}{\emph{arXiv preprint arXiv:2402.06326}} (\bibinfo{year}{2024}).
\newblock


\bibitem[\protect\citeauthoryear{Cong, Zhang, Kang, Yuan, Wu, Zhou, Tong, and Mahdavi}{Cong et~al\mbox{.}}{2023}]%
        {graphmixer}
\bibfield{author}{\bibinfo{person}{Weilin Cong}, \bibinfo{person}{Si Zhang}, \bibinfo{person}{Jian Kang}, \bibinfo{person}{Baichuan Yuan}, \bibinfo{person}{Hao Wu}, \bibinfo{person}{Xin Zhou}, \bibinfo{person}{Hanghang Tong}, {and} \bibinfo{person}{Mehrdad Mahdavi}.} \bibinfo{year}{2023}\natexlab{}.
\newblock \showarticletitle{Do We Really Need Complicated Model Architectures For Temporal Networks?}
\newblock \bibinfo{journal}{\emph{arXiv preprint arXiv:2302.11636}} (\bibinfo{year}{2023}).
\newblock


\bibitem[\protect\citeauthoryear{Feng, Li, Zhang, and Zhou}{Feng et~al\mbox{.}}{2023}]%
        {classincrease}
\bibfield{author}{\bibinfo{person}{Kaituo Feng}, \bibinfo{person}{Changsheng Li}, \bibinfo{person}{Xiaolu Zhang}, {and} \bibinfo{person}{Jun Zhou}.} \bibinfo{year}{2023}\natexlab{}.
\newblock \showarticletitle{Towards Open Temporal Graph Neural Networks}.
\newblock \bibinfo{journal}{\emph{arXiv preprint arXiv:2303.15015}} (\bibinfo{year}{2023}).
\newblock


\bibitem[\protect\citeauthoryear{Gao, Yang, Zhang, Zhou, and Hu}{Gao et~al\mbox{.}}{2021}]%
        {graphnas}
\bibfield{author}{\bibinfo{person}{Yang Gao}, \bibinfo{person}{Hong Yang}, \bibinfo{person}{Peng Zhang}, \bibinfo{person}{Chuan Zhou}, {and} \bibinfo{person}{Yue Hu}.} \bibinfo{year}{2021}\natexlab{}.
\newblock \showarticletitle{Graph neural architecture search}. In \bibinfo{booktitle}{\emph{International joint conference on artificial intelligence}}. International Joint Conference on Artificial Intelligence.
\newblock


\bibitem[\protect\citeauthoryear{Gianstefani, Longo, and Riccaboni}{Gianstefani et~al\mbox{.}}{2022}]%
        {finacialecho}
\bibfield{author}{\bibinfo{person}{Ilaria Gianstefani}, \bibinfo{person}{Luigi Longo}, {and} \bibinfo{person}{Massimo Riccaboni}.} \bibinfo{year}{2022}\natexlab{}.
\newblock \showarticletitle{The echo chamber effect resounds on financial markets: A social media alert system for meme stocks}.
\newblock \bibinfo{journal}{\emph{arXiv preprint arXiv:2203.13790}} (\bibinfo{year}{2022}).
\newblock


\bibitem[\protect\citeauthoryear{Guo, Stutz, and Schiele}{Guo et~al\mbox{.}}{2023}]%
        {attentiondiversity}
\bibfield{author}{\bibinfo{person}{Yong Guo}, \bibinfo{person}{David Stutz}, {and} \bibinfo{person}{Bernt Schiele}.} \bibinfo{year}{2023}\natexlab{}.
\newblock \showarticletitle{Robustifying token attention for vision transformers}. In \bibinfo{booktitle}{\emph{Proceedings of the IEEE/CVF International Conference on Computer Vision}}. \bibinfo{pages}{17557--17568}.
\newblock


\bibitem[\protect\citeauthoryear{Huan, Quanming, and Weiwei}{Huan et~al\mbox{.}}{2021}]%
        {same}
\bibfield{author}{\bibinfo{person}{ZHAO Huan}, \bibinfo{person}{YAO Quanming}, {and} \bibinfo{person}{TU Weiwei}.} \bibinfo{year}{2021}\natexlab{}.
\newblock \showarticletitle{Search to aggregate neighborhood for graph neural network}. In \bibinfo{booktitle}{\emph{2021 IEEE 37th International Conference on Data Engineering (ICDE)}}. IEEE, \bibinfo{pages}{552--563}.
\newblock


\bibitem[\protect\citeauthoryear{Jia, Xiong, Nan, Zhang, Zhao, and Wen}{Jia et~al\mbox{.}}{2023}]%
        {magic}
\bibfield{author}{\bibinfo{person}{Zian Jia}, \bibinfo{person}{Yun Xiong}, \bibinfo{person}{Yuhong Nan}, \bibinfo{person}{Yao Zhang}, \bibinfo{person}{Jinjing Zhao}, {and} \bibinfo{person}{Mi Wen}.} \bibinfo{year}{2023}\natexlab{}.
\newblock \showarticletitle{MAGIC: Detecting Advanced Persistent Threats via Masked Graph Representation Learning}.
\newblock \bibinfo{journal}{\emph{arXiv preprint arXiv:2310.09831}} (\bibinfo{year}{2023}).
\newblock


\bibitem[\protect\citeauthoryear{Keriven}{Keriven}{2022}]%
        {over-smoothing}
\bibfield{author}{\bibinfo{person}{Nicolas Keriven}.} \bibinfo{year}{2022}\natexlab{}.
\newblock \showarticletitle{Not too little, not too much: a theoretical analysis of graph (over) smoothing}.
\newblock \bibinfo{journal}{\emph{Advances in Neural Information Processing Systems}}  \bibinfo{volume}{35} (\bibinfo{year}{2022}), \bibinfo{pages}{2268--2281}.
\newblock


\bibitem[\protect\citeauthoryear{Kumar, Zhang, and Leskovec}{Kumar et~al\mbox{.}}{2019}]%
        {jodie}
\bibfield{author}{\bibinfo{person}{Srijan Kumar}, \bibinfo{person}{Xikun Zhang}, {and} \bibinfo{person}{Jure Leskovec}.} \bibinfo{year}{2019}\natexlab{}.
\newblock \showarticletitle{Predicting dynamic embedding trajectory in temporal interaction networks}. In \bibinfo{booktitle}{\emph{Proceedings of the 25th ACM SIGKDD international conference on knowledge discovery \& data mining}}. \bibinfo{pages}{1269--1278}.
\newblock


\bibitem[\protect\citeauthoryear{Lai, Zha, Zhou, and Hu}{Lai et~al\mbox{.}}{2020}]%
        {policygnn}
\bibfield{author}{\bibinfo{person}{Kwei-Herng Lai}, \bibinfo{person}{Daochen Zha}, \bibinfo{person}{Kaixiong Zhou}, {and} \bibinfo{person}{Xia Hu}.} \bibinfo{year}{2020}\natexlab{}.
\newblock \showarticletitle{Policy-gnn: Aggregation optimization for graph neural networks}. In \bibinfo{booktitle}{\emph{Proceedings of the 26th ACM SIGKDD International Conference on Knowledge Discovery \& Data Mining}}. \bibinfo{pages}{461--471}.
\newblock


\bibitem[\protect\citeauthoryear{Liu, Chen, Li, Zhou, Li, Song, and Qi}{Liu et~al\mbox{.}}{2019}]%
        {geniepath}
\bibfield{author}{\bibinfo{person}{Ziqi Liu}, \bibinfo{person}{Chaochao Chen}, \bibinfo{person}{Longfei Li}, \bibinfo{person}{Jun Zhou}, \bibinfo{person}{Xiaolong Li}, \bibinfo{person}{Le Song}, {and} \bibinfo{person}{Yuan Qi}.} \bibinfo{year}{2019}\natexlab{}.
\newblock \showarticletitle{Geniepath: Graph neural networks with adaptive receptive paths}. In \bibinfo{booktitle}{\emph{Proceedings of the AAAI Conference on Artificial Intelligence}}, Vol.~\bibinfo{volume}{33}. \bibinfo{pages}{4424--4431}.
\newblock


\bibitem[\protect\citeauthoryear{Luo, Gu, Zhou, Xiong, and Gao}{Luo et~al\mbox{.}}{2023}]%
        {luo}
\bibfield{author}{\bibinfo{person}{Yang Luo}, \bibinfo{person}{Zehao Gu}, \bibinfo{person}{Shiyang Zhou}, \bibinfo{person}{Yun Xiong}, {and} \bibinfo{person}{Xiaofeng Gao}.} \bibinfo{year}{2023}\natexlab{}.
\newblock \showarticletitle{Meteorology-Assisted Spatio-Temporal Graph Network for Uncivilized Urban Event Prediction}. In \bibinfo{booktitle}{\emph{2023 IEEE International Conference on Data Mining (ICDM)}}. IEEE, \bibinfo{pages}{468--477}.
\newblock


\bibitem[\protect\citeauthoryear{McGinn, Birch, Akroyd, Molina-Solana, Guo, and Knottenbelt}{McGinn et~al\mbox{.}}{2016}]%
        {finacialimpact2}
\bibfield{author}{\bibinfo{person}{Dan McGinn}, \bibinfo{person}{David Birch}, \bibinfo{person}{David Akroyd}, \bibinfo{person}{Miguel Molina-Solana}, \bibinfo{person}{Yike Guo}, {and} \bibinfo{person}{William~J Knottenbelt}.} \bibinfo{year}{2016}\natexlab{}.
\newblock \showarticletitle{Visualizing dynamic bitcoin transaction patterns}.
\newblock \bibinfo{journal}{\emph{Big data}} \bibinfo{volume}{4}, \bibinfo{number}{2} (\bibinfo{year}{2016}), \bibinfo{pages}{109--119}.
\newblock


\bibitem[\protect\citeauthoryear{Poursafaei, Huang, Pelrine, and Rabbany}{Poursafaei et~al\mbox{.}}{2022}]%
        {edgebank}
\bibfield{author}{\bibinfo{person}{Farimah Poursafaei}, \bibinfo{person}{Shenyang Huang}, \bibinfo{person}{Kellin Pelrine}, {and} \bibinfo{person}{Reihaneh Rabbany}.} \bibinfo{year}{2022}\natexlab{}.
\newblock \showarticletitle{Towards better evaluation for dynamic link prediction}.
\newblock \bibinfo{journal}{\emph{Advances in Neural Information Processing Systems}}  \bibinfo{volume}{35} (\bibinfo{year}{2022}), \bibinfo{pages}{32928--32941}.
\newblock


\bibitem[\protect\citeauthoryear{Rossi, Chamberlain, Frasca, Eynard, Monti, and Bronstein}{Rossi et~al\mbox{.}}{2020}]%
        {tgn}
\bibfield{author}{\bibinfo{person}{Emanuele Rossi}, \bibinfo{person}{Ben Chamberlain}, \bibinfo{person}{Fabrizio Frasca}, \bibinfo{person}{Davide Eynard}, \bibinfo{person}{Federico Monti}, {and} \bibinfo{person}{Michael Bronstein}.} \bibinfo{year}{2020}\natexlab{}.
\newblock \showarticletitle{Temporal graph networks for deep learning on dynamic graphs}.
\newblock \bibinfo{journal}{\emph{arXiv preprint arXiv:2006.10637}} (\bibinfo{year}{2020}).
\newblock


\bibitem[\protect\citeauthoryear{Souza, Mesquita, Kaski, and Garg}{Souza et~al\mbox{.}}{2022}]%
        {pint}
\bibfield{author}{\bibinfo{person}{Amauri Souza}, \bibinfo{person}{Diego Mesquita}, \bibinfo{person}{Samuel Kaski}, {and} \bibinfo{person}{Vikas Garg}.} \bibinfo{year}{2022}\natexlab{}.
\newblock \showarticletitle{Provably expressive temporal graph networks}.
\newblock \bibinfo{journal}{\emph{Advances in Neural Information Processing Systems}}  \bibinfo{volume}{35} (\bibinfo{year}{2022}), \bibinfo{pages}{32257--32269}.
\newblock


\bibitem[\protect\citeauthoryear{Sunstein}{Sunstein}{2006}]%
        {informationcoco}
\bibfield{author}{\bibinfo{person}{Cass~R Sunstein}.} \bibinfo{year}{2006}\natexlab{}.
\newblock \bibinfo{booktitle}{\emph{Infotopia: How many minds produce knowledge}}.
\newblock \bibinfo{publisher}{Oxford University Press}.
\newblock


\bibitem[\protect\citeauthoryear{Tan, Zhang, Liu, Zha, Li, Chen, Choi, and Hu}{Tan et~al\mbox{.}}{2023}]%
        {subgraph}
\bibfield{author}{\bibinfo{person}{Qiaoyu Tan}, \bibinfo{person}{Xin Zhang}, \bibinfo{person}{Ninghao Liu}, \bibinfo{person}{Daochen Zha}, \bibinfo{person}{Li Li}, \bibinfo{person}{Rui Chen}, \bibinfo{person}{Soo-Hyun Choi}, {and} \bibinfo{person}{Xia Hu}.} \bibinfo{year}{2023}\natexlab{}.
\newblock \showarticletitle{Bring your own view: Graph neural networks for link prediction with personalized subgraph selection}. In \bibinfo{booktitle}{\emph{Proceedings of the Sixteenth ACM International Conference on Web Search and Data Mining}}. \bibinfo{pages}{625--633}.
\newblock


\bibitem[\protect\citeauthoryear{Tang, Yao, Sun, Wang, Tang, Aggarwal, Mitra, and Wang}{Tang et~al\mbox{.}}{2020}]%
        {normalize1}
\bibfield{author}{\bibinfo{person}{Xianfeng Tang}, \bibinfo{person}{Huaxiu Yao}, \bibinfo{person}{Yiwei Sun}, \bibinfo{person}{Yiqi Wang}, \bibinfo{person}{Jiliang Tang}, \bibinfo{person}{Charu Aggarwal}, \bibinfo{person}{Prasenjit Mitra}, {and} \bibinfo{person}{Suhang Wang}.} \bibinfo{year}{2020}\natexlab{}.
\newblock \showarticletitle{Investigating and mitigating degree-related biases in graph convoltuional networks}. In \bibinfo{booktitle}{\emph{Proceedings of the 29th ACM International Conference on Information \& Knowledge Management}}. \bibinfo{pages}{1435--1444}.
\newblock


\bibitem[\protect\citeauthoryear{Trivedi, Farajtabar, Biswal, and Zha}{Trivedi et~al\mbox{.}}{2019}]%
        {dyrep}
\bibfield{author}{\bibinfo{person}{Rakshit Trivedi}, \bibinfo{person}{Mehrdad Farajtabar}, \bibinfo{person}{Prasenjeet Biswal}, {and} \bibinfo{person}{Hongyuan Zha}.} \bibinfo{year}{2019}\natexlab{}.
\newblock \showarticletitle{Dyrep: Learning representations over dynamic graphs}. In \bibinfo{booktitle}{\emph{International conference on learning representations}}.
\newblock


\bibitem[\protect\citeauthoryear{Vaswani, Shazeer, Parmar, Uszkoreit, Jones, Gomez, Kaiser, and Polosukhin}{Vaswani et~al\mbox{.}}{2017}]%
        {transformer}
\bibfield{author}{\bibinfo{person}{Ashish Vaswani}, \bibinfo{person}{Noam Shazeer}, \bibinfo{person}{Niki Parmar}, \bibinfo{person}{Jakob Uszkoreit}, \bibinfo{person}{Llion Jones}, \bibinfo{person}{Aidan~N Gomez}, \bibinfo{person}{{\L}ukasz Kaiser}, {and} \bibinfo{person}{Illia Polosukhin}.} \bibinfo{year}{2017}\natexlab{}.
\newblock \showarticletitle{Attention is all you need}.
\newblock \bibinfo{journal}{\emph{Advances in neural information processing systems}}  \bibinfo{volume}{30} (\bibinfo{year}{2017}).
\newblock


\bibitem[\protect\citeauthoryear{Wang, Cai, Liang, Ding, Wang, Bhatia, and Hooi}{Wang et~al\mbox{.}}{2021a}]%
        {mata}
\bibfield{author}{\bibinfo{person}{Yiwei Wang}, \bibinfo{person}{Yujun Cai}, \bibinfo{person}{Yuxuan Liang}, \bibinfo{person}{Henghui Ding}, \bibinfo{person}{Changhu Wang}, \bibinfo{person}{Siddharth Bhatia}, {and} \bibinfo{person}{Bryan Hooi}.} \bibinfo{year}{2021}\natexlab{a}.
\newblock \showarticletitle{Adaptive data augmentation on temporal graphs}.
\newblock \bibinfo{journal}{\emph{Advances in Neural Information Processing Systems}}  \bibinfo{volume}{34} (\bibinfo{year}{2021}), \bibinfo{pages}{1440--1452}.
\newblock


\bibitem[\protect\citeauthoryear{Wang, Chang, Liu, Leskovec, and Li}{Wang et~al\mbox{.}}{2021b}]%
        {caw}
\bibfield{author}{\bibinfo{person}{Yanbang Wang}, \bibinfo{person}{Yen-Yu Chang}, \bibinfo{person}{Yunyu Liu}, \bibinfo{person}{Jure Leskovec}, {and} \bibinfo{person}{Pan Li}.} \bibinfo{year}{2021}\natexlab{b}.
\newblock \showarticletitle{Inductive representation learning in temporal networks via causal anonymous walks}.
\newblock \bibinfo{journal}{\emph{arXiv preprint arXiv:2101.05974}} (\bibinfo{year}{2021}).
\newblock


\bibitem[\protect\citeauthoryear{Wang, Di, and Chen}{Wang et~al\mbox{.}}{2021c}]%
        {autogel}
\bibfield{author}{\bibinfo{person}{Zhili Wang}, \bibinfo{person}{Shimin Di}, {and} \bibinfo{person}{Lei Chen}.} \bibinfo{year}{2021}\natexlab{c}.
\newblock \showarticletitle{Autogel: An automated graph neural network with explicit link information}.
\newblock \bibinfo{journal}{\emph{Advances in Neural Information Processing Systems}}  \bibinfo{volume}{34} (\bibinfo{year}{2021}), \bibinfo{pages}{24509--24522}.
\newblock


\bibitem[\protect\citeauthoryear{Wang, Di, and Chen}{Wang et~al\mbox{.}}{2023a}]%
        {messagepassing}
\bibfield{author}{\bibinfo{person}{Zhili Wang}, \bibinfo{person}{Shimin Di}, {and} \bibinfo{person}{Lei Chen}.} \bibinfo{year}{2023}\natexlab{a}.
\newblock \showarticletitle{A Message Passing Neural Network Space for Better Capturing Data-dependent Receptive Fields}. In \bibinfo{booktitle}{\emph{Proceedings of the 29th ACM SIGKDD Conference on Knowledge Discovery and Data Mining}}. \bibinfo{pages}{2489--2501}.
\newblock


\bibitem[\protect\citeauthoryear{Wang, Di, and Chen}{Wang et~al\mbox{.}}{2023b}]%
        {dense}
\bibfield{author}{\bibinfo{person}{Zhili Wang}, \bibinfo{person}{Shimin Di}, {and} \bibinfo{person}{Lei Chen}.} \bibinfo{year}{2023}\natexlab{b}.
\newblock \showarticletitle{A Message Passing Neural Network Space for Better Capturing Data-dependent Receptive Fields}. In \bibinfo{booktitle}{\emph{Proceedings of the 29th ACM SIGKDD Conference on Knowledge Discovery and Data Mining}}. \bibinfo{pages}{2489--2501}.
\newblock


\bibitem[\protect\citeauthoryear{Wei, Zhao, Yao, and He}{Wei et~al\mbox{.}}{2021}]%
        {pooling}
\bibfield{author}{\bibinfo{person}{Lanning Wei}, \bibinfo{person}{Huan Zhao}, \bibinfo{person}{Quanming Yao}, {and} \bibinfo{person}{Zhiqiang He}.} \bibinfo{year}{2021}\natexlab{}.
\newblock \showarticletitle{Pooling architecture search for graph classification}. In \bibinfo{booktitle}{\emph{Proceedings of the 30th ACM International Conference on Information \& Knowledge Management}}. \bibinfo{pages}{2091--2100}.
\newblock


\bibitem[\protect\citeauthoryear{Xiang, Zhu, Cheng, Li, Zhao, Ouyang, Chen, and Zheng}{Xiang et~al\mbox{.}}{2023}]%
        {noise}
\bibfield{author}{\bibinfo{person}{Sheng Xiang}, \bibinfo{person}{Mingzhi Zhu}, \bibinfo{person}{Dawei Cheng}, \bibinfo{person}{Enxia Li}, \bibinfo{person}{Ruihui Zhao}, \bibinfo{person}{Yi Ouyang}, \bibinfo{person}{Ling Chen}, {and} \bibinfo{person}{Yefeng Zheng}.} \bibinfo{year}{2023}\natexlab{}.
\newblock \showarticletitle{Semi-supervised credit card fraud detection via attribute-driven graph representation}. In \bibinfo{booktitle}{\emph{Proceedings of the AAAI Conference on Artificial Intelligence}}, Vol.~\bibinfo{volume}{37}. \bibinfo{pages}{14557--14565}.
\newblock


\bibitem[\protect\citeauthoryear{Xu, Ruan, Korpeoglu, Kumar, and Achan}{Xu et~al\mbox{.}}{2020b}]%
        {tgat}
\bibfield{author}{\bibinfo{person}{Da Xu}, \bibinfo{person}{Chuanwei Ruan}, \bibinfo{person}{Evren Korpeoglu}, \bibinfo{person}{Sushant Kumar}, {and} \bibinfo{person}{Kannan Achan}.} \bibinfo{year}{2020}\natexlab{b}.
\newblock \showarticletitle{Inductive representation learning on temporal graphs}.
\newblock \bibinfo{journal}{\emph{arXiv preprint arXiv:2002.07962}} (\bibinfo{year}{2020}).
\newblock


\bibitem[\protect\citeauthoryear{Xu, Chen, Li, and Wang}{Xu et~al\mbox{.}}{2020a}]%
        {informationcoco1}
\bibfield{author}{\bibinfo{person}{Huimin Xu}, \bibinfo{person}{Zhicong Chen}, \bibinfo{person}{Ruiqi Li}, {and} \bibinfo{person}{Cheng-Jun Wang}.} \bibinfo{year}{2020}\natexlab{a}.
\newblock \showarticletitle{The geometry of information cocoon: Analyzing the cultural space with word embedding models}.
\newblock \bibinfo{journal}{\emph{arXiv preprint arXiv:2007.10083}} (\bibinfo{year}{2020}).
\newblock


\bibitem[\protect\citeauthoryear{Xu, Li, Tian, Sonobe, Kawarabayashi, and Jegelka}{Xu et~al\mbox{.}}{2018}]%
        {jump}
\bibfield{author}{\bibinfo{person}{Keyulu Xu}, \bibinfo{person}{Chengtao Li}, \bibinfo{person}{Yonglong Tian}, \bibinfo{person}{Tomohiro Sonobe}, \bibinfo{person}{Ken-ichi Kawarabayashi}, {and} \bibinfo{person}{Stefanie Jegelka}.} \bibinfo{year}{2018}\natexlab{}.
\newblock \showarticletitle{Representation learning on graphs with jumping knowledge networks}. In \bibinfo{booktitle}{\emph{International conference on machine learning}}. PMLR, \bibinfo{pages}{5453--5462}.
\newblock


\bibitem[\protect\citeauthoryear{Ying, Cai, Luo, Zheng, Ke, He, Shen, and Liu}{Ying et~al\mbox{.}}{2021}]%
        {graformer}
\bibfield{author}{\bibinfo{person}{Chengxuan Ying}, \bibinfo{person}{Tianle Cai}, \bibinfo{person}{Shengjie Luo}, \bibinfo{person}{Shuxin Zheng}, \bibinfo{person}{Guolin Ke}, \bibinfo{person}{Di He}, \bibinfo{person}{Yanming Shen}, {and} \bibinfo{person}{Tie-Yan Liu}.} \bibinfo{year}{2021}\natexlab{}.
\newblock \showarticletitle{Do transformers really perform badly for graph representation?}
\newblock \bibinfo{journal}{\emph{Advances in Neural Information Processing Systems}}  \bibinfo{volume}{34} (\bibinfo{year}{2021}), \bibinfo{pages}{28877--28888}.
\newblock


\bibitem[\protect\citeauthoryear{Yu, Sun, Du, and Lv}{Yu et~al\mbox{.}}{2023}]%
        {dygformer}
\bibfield{author}{\bibinfo{person}{Le Yu}, \bibinfo{person}{Leilei Sun}, \bibinfo{person}{Bowen Du}, {and} \bibinfo{person}{Weifeng Lv}.} \bibinfo{year}{2023}\natexlab{}.
\newblock \showarticletitle{Towards Better Dynamic Graph Learning: New Architecture and Unified Library}.
\newblock \bibinfo{journal}{\emph{arXiv preprint arXiv:2303.13047}} (\bibinfo{year}{2023}).
\newblock


\bibitem[\protect\citeauthoryear{Yu, Si, Hu, and Zhang}{Yu et~al\mbox{.}}{2019}]%
        {LSTM}
\bibfield{author}{\bibinfo{person}{Yong Yu}, \bibinfo{person}{Xiaosheng Si}, \bibinfo{person}{Changhua Hu}, {and} \bibinfo{person}{Jianxun Zhang}.} \bibinfo{year}{2019}\natexlab{}.
\newblock \showarticletitle{A review of recurrent neural networks: LSTM cells and network architectures}.
\newblock \bibinfo{journal}{\emph{Neural computation}} \bibinfo{volume}{31}, \bibinfo{number}{7} (\bibinfo{year}{2019}), \bibinfo{pages}{1235--1270}.
\newblock


\bibitem[\protect\citeauthoryear{Zhang, Ren, and Urtasun}{Zhang et~al\mbox{.}}{2018}]%
        {hypernetworks}
\bibfield{author}{\bibinfo{person}{Chris Zhang}, \bibinfo{person}{Mengye Ren}, {and} \bibinfo{person}{Raquel Urtasun}.} \bibinfo{year}{2018}\natexlab{}.
\newblock \showarticletitle{Graph hypernetworks for neural architecture search}.
\newblock \bibinfo{journal}{\emph{arXiv preprint arXiv:1810.05749}} (\bibinfo{year}{2018}).
\newblock


\bibitem[\protect\citeauthoryear{Zhang, Xiong, Zhang, Sun, Chen, Jiao, and Zhu}{Zhang et~al\mbox{.}}{2023b}]%
        {rdgsl}
\bibfield{author}{\bibinfo{person}{Siwei Zhang}, \bibinfo{person}{Yun Xiong}, \bibinfo{person}{Yao Zhang}, \bibinfo{person}{Yiheng Sun}, \bibinfo{person}{Xi Chen}, \bibinfo{person}{Yizhu Jiao}, {and} \bibinfo{person}{Yangyong Zhu}.} \bibinfo{year}{2023}\natexlab{b}.
\newblock \showarticletitle{RDGSL: Dynamic Graph Representation Learning with Structure Learning}. In \bibinfo{booktitle}{\emph{Proceedings of the 32nd ACM International Conference on Information and Knowledge Management}}. \bibinfo{pages}{3174--3183}.
\newblock


\bibitem[\protect\citeauthoryear{Zhang, Xiong, Zhang, Wu, Sun, and Zhang}{Zhang et~al\mbox{.}}{2023c}]%
        {ilore}
\bibfield{author}{\bibinfo{person}{Siwei Zhang}, \bibinfo{person}{Yun Xiong}, \bibinfo{person}{Yao Zhang}, \bibinfo{person}{Xixi Wu}, \bibinfo{person}{Yiheng Sun}, {and} \bibinfo{person}{Jiawei Zhang}.} \bibinfo{year}{2023}\natexlab{c}.
\newblock \showarticletitle{iLoRE: Dynamic Graph Representation with Instant Long-term Modeling and Re-occurrence Preservation}. In \bibinfo{booktitle}{\emph{Proceedings of the 32nd ACM International Conference on Information and Knowledge Management}}. \bibinfo{pages}{3216--3225}.
\newblock


\bibitem[\protect\citeauthoryear{Zhang, Wang, Zhao, and Tang}{Zhang et~al\mbox{.}}{2012}]%
        {normalized2}
\bibfield{author}{\bibinfo{person}{Wei Zhang}, \bibinfo{person}{Xiaogang Wang}, \bibinfo{person}{Deli Zhao}, {and} \bibinfo{person}{Xiaoou Tang}.} \bibinfo{year}{2012}\natexlab{}.
\newblock \showarticletitle{Graph degree linkage: Agglomerative clustering on a directed graph}. In \bibinfo{booktitle}{\emph{Computer Vision--ECCV 2012: 12th European Conference on Computer Vision, Florence, Italy, October 7-13, 2012, Proceedings, Part I 12}}. Springer, \bibinfo{pages}{428--441}.
\newblock


\bibitem[\protect\citeauthoryear{Zhang, Xiong, Li, Shan, Ren, and Zhu}{Zhang et~al\mbox{.}}{2021}]%
        {cope}
\bibfield{author}{\bibinfo{person}{Yao Zhang}, \bibinfo{person}{Yun Xiong}, \bibinfo{person}{Dongsheng Li}, \bibinfo{person}{Caihua Shan}, \bibinfo{person}{Kan Ren}, {and} \bibinfo{person}{Yangyong Zhu}.} \bibinfo{year}{2021}\natexlab{}.
\newblock \showarticletitle{CoPE: Modeling Continuous Propagation and Evolution on Interaction Graph}. In \bibinfo{booktitle}{\emph{Proceedings of the 30th ACM International Conference on Information \& Knowledge Management}}. \bibinfo{pages}{2627--2636}.
\newblock


\bibitem[\protect\citeauthoryear{Zhang, Xiong, Liao, Sun, Jin, Zheng, and Zhu}{Zhang et~al\mbox{.}}{2023a}]%
        {tiger}
\bibfield{author}{\bibinfo{person}{Yao Zhang}, \bibinfo{person}{Yun Xiong}, \bibinfo{person}{Yongxiang Liao}, \bibinfo{person}{Yiheng Sun}, \bibinfo{person}{Yucheng Jin}, \bibinfo{person}{Xuehao Zheng}, {and} \bibinfo{person}{Yangyong Zhu}.} \bibinfo{year}{2023}\natexlab{a}.
\newblock \showarticletitle{TIGER: Temporal Interaction Graph Embedding with Restarts}. In \bibinfo{booktitle}{\emph{ACM Web Conference}}.
\newblock


\bibitem[\protect\citeauthoryear{Zhou, Endendijk, and Botzen}{Zhou et~al\mbox{.}}{2023}]%
        {finacialimpact}
\bibfield{author}{\bibinfo{person}{Fujin Zhou}, \bibinfo{person}{Thijs Endendijk}, {and} \bibinfo{person}{WJ~Wouter Botzen}.} \bibinfo{year}{2023}\natexlab{}.
\newblock \showarticletitle{A review of the financial sector impacts of risks associated with climate change}.
\newblock \bibinfo{journal}{\emph{Annual Review of Resource Economics}}  \bibinfo{volume}{15} (\bibinfo{year}{2023}), \bibinfo{pages}{233--256}.
\newblock


\bibitem[\protect\citeauthoryear{Zhou, Huang, Song, Chen, and Hu}{Zhou et~al\mbox{.}}{2022}]%
        {autognn}
\bibfield{author}{\bibinfo{person}{Kaixiong Zhou}, \bibinfo{person}{Xiao Huang}, \bibinfo{person}{Qingquan Song}, \bibinfo{person}{Rui Chen}, {and} \bibinfo{person}{Xia Hu}.} \bibinfo{year}{2022}\natexlab{}.
\newblock \showarticletitle{Auto-gnn: Neural architecture search of graph neural networks}.
\newblock \bibinfo{journal}{\emph{Frontiers in big Data}}  \bibinfo{volume}{5} (\bibinfo{year}{2022}), \bibinfo{pages}{1029307}.
\newblock


\bibitem[\protect\citeauthoryear{Zhu, Yan, Zhao, Heimann, Akoglu, and Koutra}{Zhu et~al\mbox{.}}{2020}]%
        {beyond}
\bibfield{author}{\bibinfo{person}{Jiong Zhu}, \bibinfo{person}{Yujun Yan}, \bibinfo{person}{Lingxiao Zhao}, \bibinfo{person}{Mark Heimann}, \bibinfo{person}{Leman Akoglu}, {and} \bibinfo{person}{Danai Koutra}.} \bibinfo{year}{2020}\natexlab{}.
\newblock \showarticletitle{Beyond homophily in graph neural networks: Current limitations and effective designs}.
\newblock \bibinfo{journal}{\emph{Advances in neural information processing systems}}  \bibinfo{volume}{33} (\bibinfo{year}{2020}), \bibinfo{pages}{7793--7804}.
\newblock


\end{thebibliography}


\clearpage
\newpage
\appendix
\section{appendix}
\subsection{Notations}
We summarize relevant notations and their descriptions in this paper as shown in Table \ref{notations}.

\begin{table}[h]
  \caption{Important notations in this paper.}
  \label{notations}
  \small
  \begin{tabular}{cc}
    \toprule
    Notations & Description or definition  \\
    \midrule 
    $\mathbf{f}_i(t)$  & Occurrence frequency feature of node $i$'s neighbors at $t$\\
    $\mathbf{R}_i(t)$ & Occurrence encoding of node $i$'s neighbors at $t$ \\
    $\hat{a}_{ij}^{(k)}$ & Enhanced attention from nodes $i$ to $j$ in the $k$-th layer \\
    $u^{(k)}$ & The state of the $k$-th layer\\
    $\mathbf{c}_{*}^{(k)}$ & Outdated-decay cell state in the $k$-th layer \\
    \midrule
    $\mathcal{L}_{\text{NDP}}$ & Loss function of neighbor diversity penalty \\
    $\mathcal{L}_{\text{link}}$ & Loss function of temporal link prediction \\
  \bottomrule
\end{tabular}
\end{table}

\subsection{Datasets}
We employ five datasets, including four public datasets and one dataset introduced in our paper.

The four public datasets%
\footnote{\url{https://snap.stanford.edu/jodie/}} used in our paper are provided by JODIE \cite{jodie}. (i) Wikipedia is a temporal interaction graph capturing edits made by users on Wikipedia pages. (ii) Reddit is a temporal interaction graph recording user posts in different subreddits. In both Wikipedia and Reddit, the edge feature dimension is 172, and user nodes are dynamically labeled to indicate whether they have been banned. (iii) MOOC is a temporal interaction graph that tracks interactions between students and online courses, with dynamic labels on nodes indicating whether students drop out of a course. (iv) LastFM is a temporal interaction graph that logs events between users and songs, but it does not include dynamic labels.

Most existing temporal interaction graph datasets are primarily focused on social or event networks. However, interaction data in the financial domain differs significantly. It typically involves a considerably larger number of participants, resulting in sparser networks. Consequently, learning tasks on such networks tend to be more challenging compared to those on existing available networks. This increased difficulty provides a more rigorous testbed for validating the capabilities of various TGNs, making a more effective benchmark to assess their performance and robustness.

In this paper, we release \textbf{TemFin}, a new temporal interaction graph benchmark dataset that records the financial transfer transactions between bank accounts. Within TemFin, bank accounts are represented as nodes, and the financial transactions with timestamp information occurring between two accounts are modeled as timestamped interactions. Additionally, the information about the transactions, \eg, amount and fund channel, is converted into a 154-dimensional vector through one-hot encoding, which then serves as the edge feature for each transaction. Furthermore, the source account nodes are assigned dynamic labels, indicating whether the corresponding account is implicated in money laundering activity. In the TemFin dataset, the Temporal Link Prediction task is designed to forecast whether one account will transfer funds to another at a given future time. Meanwhile, the Evolving Node Classification task concentrates on determining whether a source account in a specific transaction is implicated in money laundering activity. The detailed statistics of all datasets are summarized in Table \ref{datasets}.

\subsection{Implementation Details}
\label{baselines}
For baselines, all baselines employ the same experimental settings as TIGE \cite{tiger}, thus ensuring comparability in our evaluation process. The detailed default hyper-parameters can be found in its publication. 

For our {\model}, we integrate it into three representative TGNs, \ie, TGN \cite{tgn}, TIGE \cite{tiger}, and DyGFormer \cite{dygformer}, chosen for their superior performance and widespread recognition. To guarantee fairness in our evaluation, all hyper-parameters in these three backbones remain consistent with their original implementations. We implement {\model} in PyTorch based on their corresponding official implementations. Unless otherwise stated, we use the default hyper-parameters of {\model} summarized in Table \ref{tab:defalthyperparameters}. 

Experiments on all datasets are conducted on a single server with 72 cores, 32GB memory, and one Nvidia Tesla V100 GPU.
\begin{table}[h]
\centering
\caption{Default hyper-parameters of SEAN.}
\label{tab:defalthyperparameters}
\begin{tabular}{cc}
\toprule
\textbf{Hyper-parameter} & \textbf{Value} \\ 
\hline
Batch size               & 200            \\ 
Learning rate            & 0.0001         \\ 
Optimizer                & Adam           \\ 
Number of layers         & 1              \\ 
Number of attention heads & 2             \\ 
Threshold $\tau$         & 0.1            \\ 
\bottomrule
\end{tabular}
\end{table}

\subsection{Limitations}
We consider two main limitations for our {\model}:
\begin{itemize}[leftmargin=*]
    \item While {\model} enhances model performance, it inevitably introduces some complexity, which may be unavoidable.
    \item Our proposed adaptive pruning module serves as a residual link in the aggregation process and improves the overall model performance. However, when the neighborhood is extremely sparse, the pruning mechanism might lose some useful information.
\end{itemize}
We will address these limitations in our future work by exploring more efficient adaptive neighborhood encoding methods in temporal interaction graph modeling.

\begin{table*}
  \caption{Detailed statistics of datasets.}
  \label{datasets}
  \begin{tabular}{ccccccccc}
    \toprule
    Datasets & Domain & \# Nodes & \# Edges & \# Edge Feature & Duration & Label & \% Positive Labels \\
    \midrule
    Wikipedia & social & 9,227 & 157,474 & 172 & 1 month &editing ban & 0.14\%\\
    Reddit & social & 10,984 & 672,447 &172 & 1 month &posting ban & 0.05\% \\
    MOOC & event & 7,144 & 411,749 &0 & 17 month &course dropout & 0.98\% \\
    LastFM & event & 1,980 & 1,293,103 & 0 & 1 month & --- & ---\\
    TemFin (\textit{new}) & finance & 33,245 & 709,774 & 154 & half a month & money laundering & 0.71\%\\
  \bottomrule
\end{tabular}
\end{table*}

\begin{figure*}
    \centering
    \includegraphics[width=\linewidth]{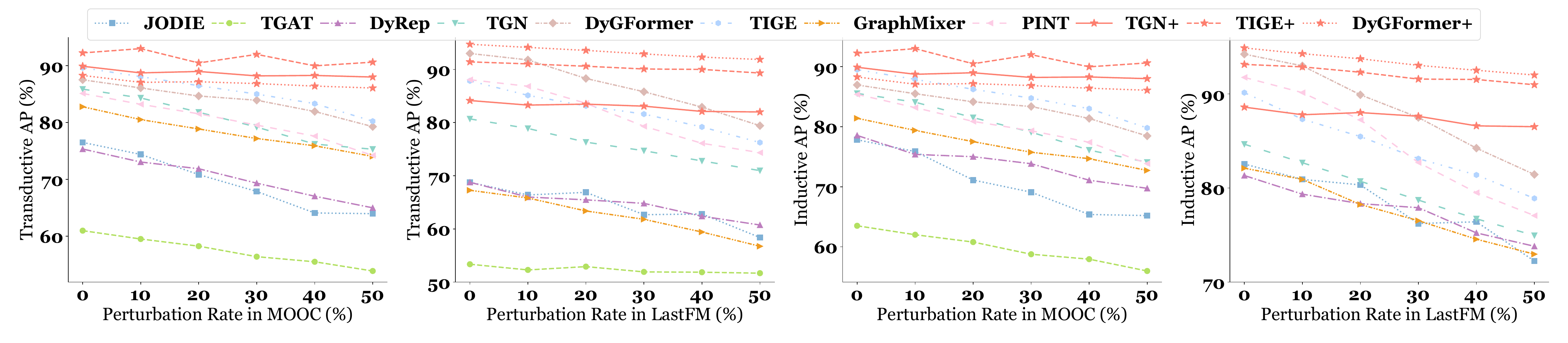}
    \captionsetup{justification=centering}
    \caption{Robustness to noisy neighborhoods on MOOC and LastFM in different perturbation rates. \\ (Supplementary results for Section \ref{sec:noisydata})}
    \label{fig:app_noise_mooc_lastfm}
\end{figure*}

\begin{figure}
    \includegraphics[width=\linewidth]{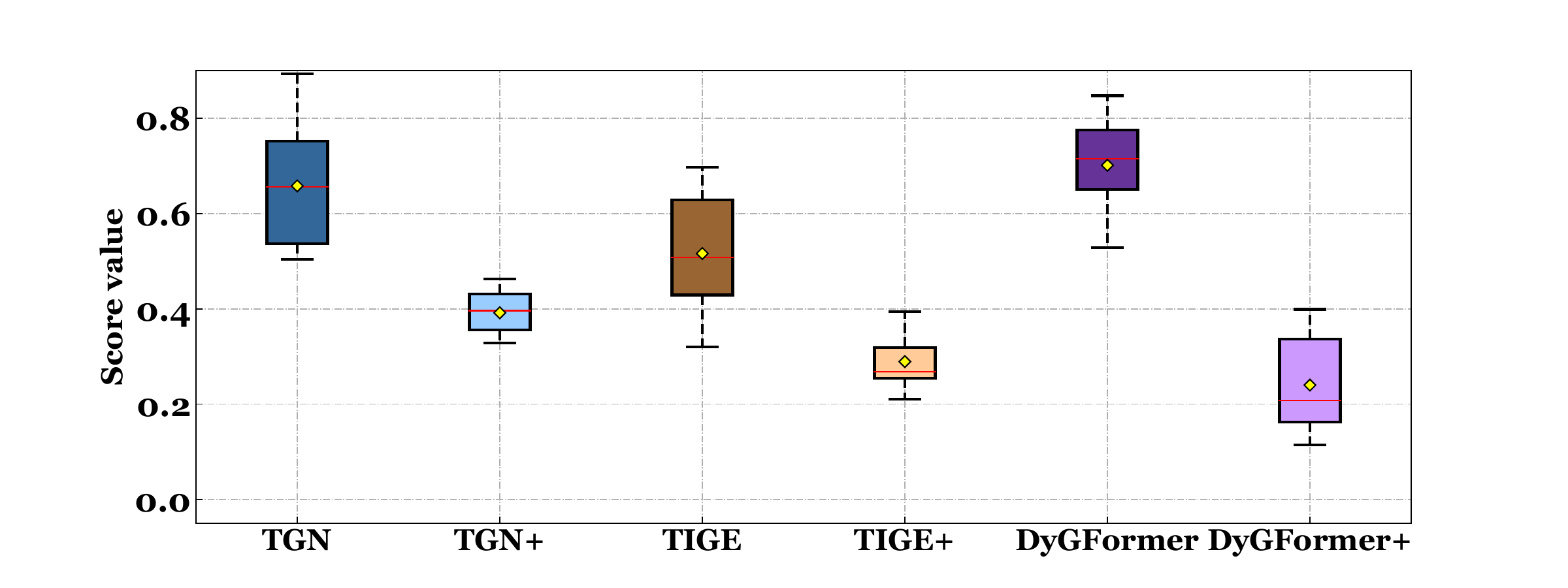}
    \captionsetup{justification=centering}
    \caption{Attention scores assigned from randomly selected nodes to their perturbed neighbors on noisy Reddit.  \\ (Supplementary results for Section \ref{sec:noisydata})}
    \label{fig:app_noise_reddit_score}
\end{figure}

\begin{figure}
  \centering
  \begin{subfigure}{.49\linewidth}
    \centering
    \includegraphics[width=0.49\linewidth]{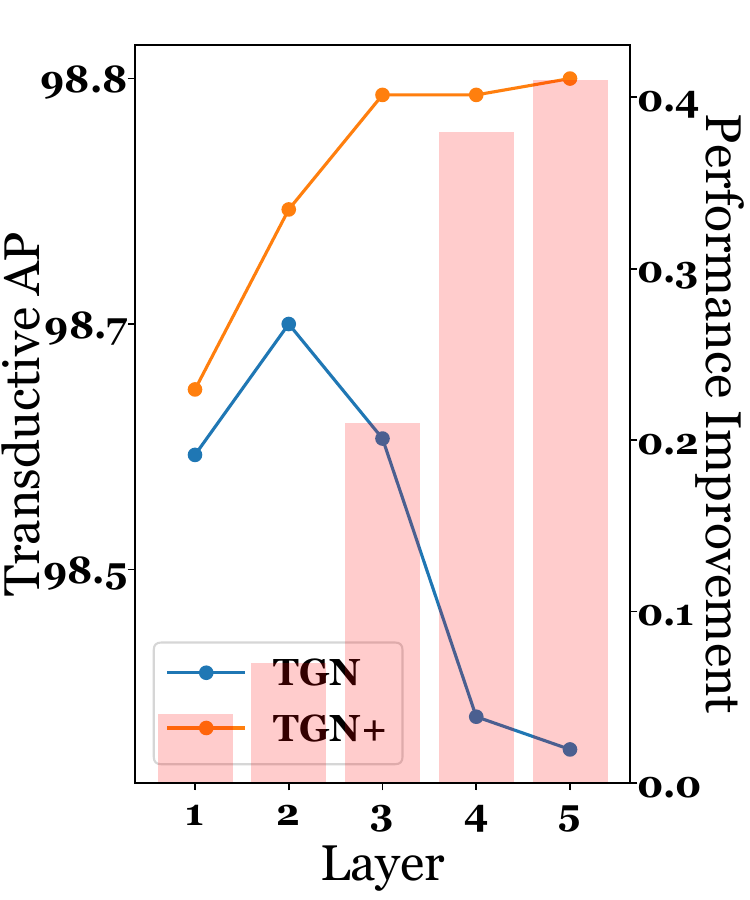}
    \includegraphics[width=0.49\linewidth]{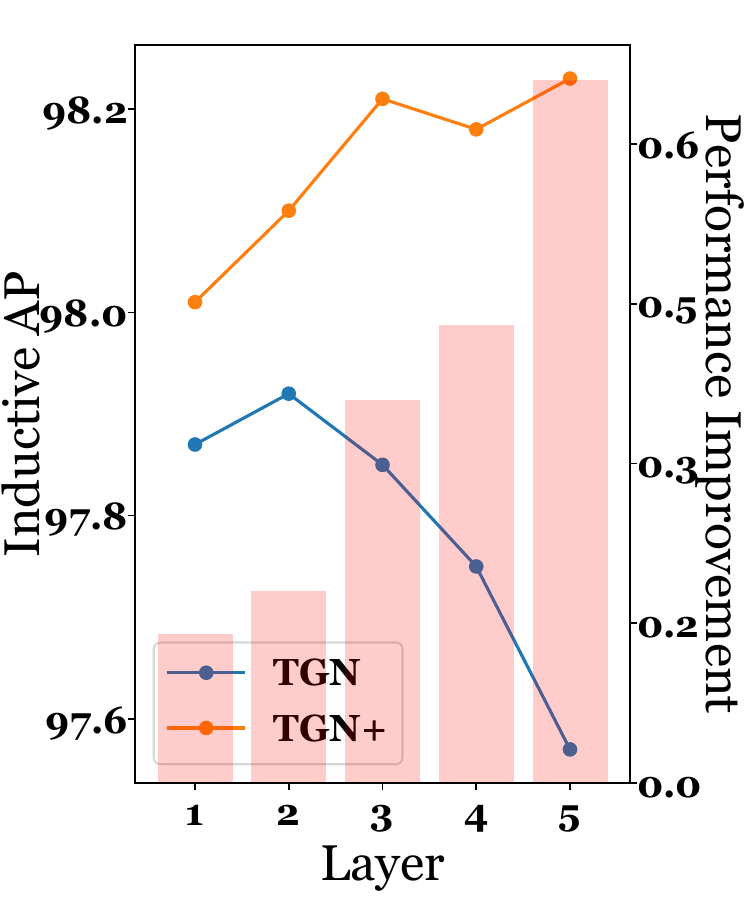}
    \caption{TGN model.}
  \end{subfigure}
  \centering
  \begin{subfigure}{.49\linewidth}
    \centering
    \includegraphics[width=.49\linewidth]{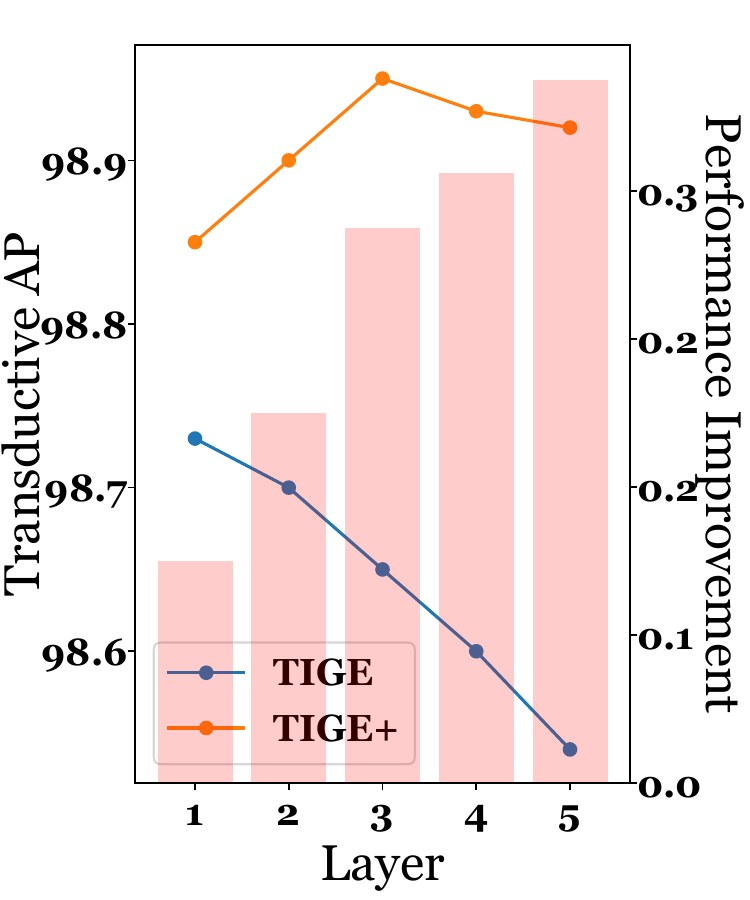}
    \includegraphics[width=.49\linewidth]{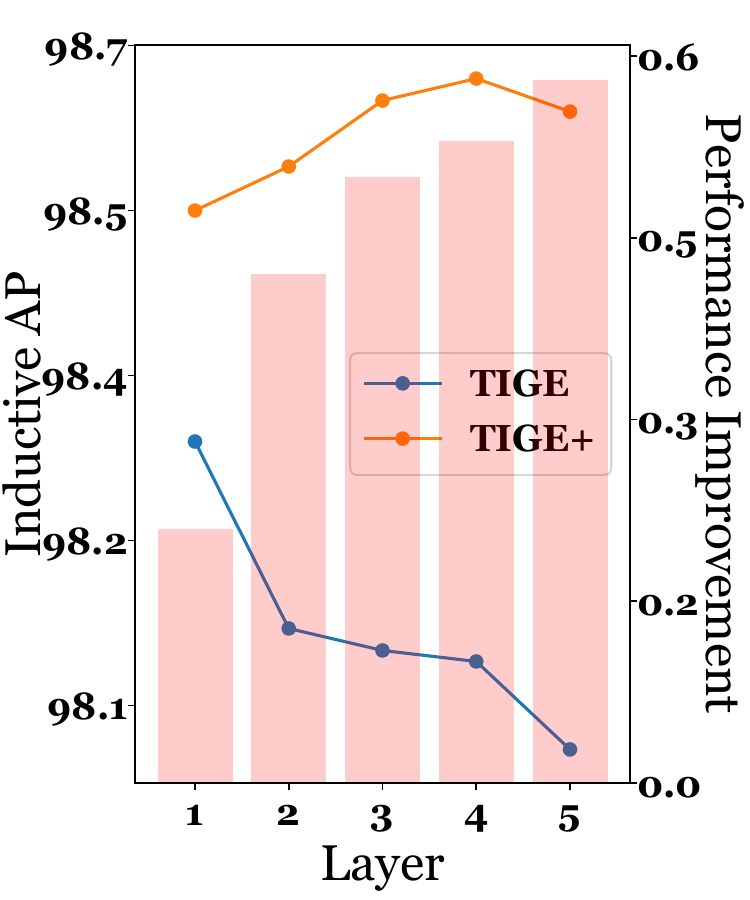}
    \caption{TIGE model.}
  \end{subfigure}
  \captionsetup{justification=centering}
  \caption{Robustness to expanded neighborhoods on Wikipedia with TGN and TIGE models.  \\ (Supplementary results for Section \ref{sec:layer})}
  \label{fig:app_layer}
\end{figure}

\begin{figure}
    \centering
    \includegraphics[width=\linewidth]{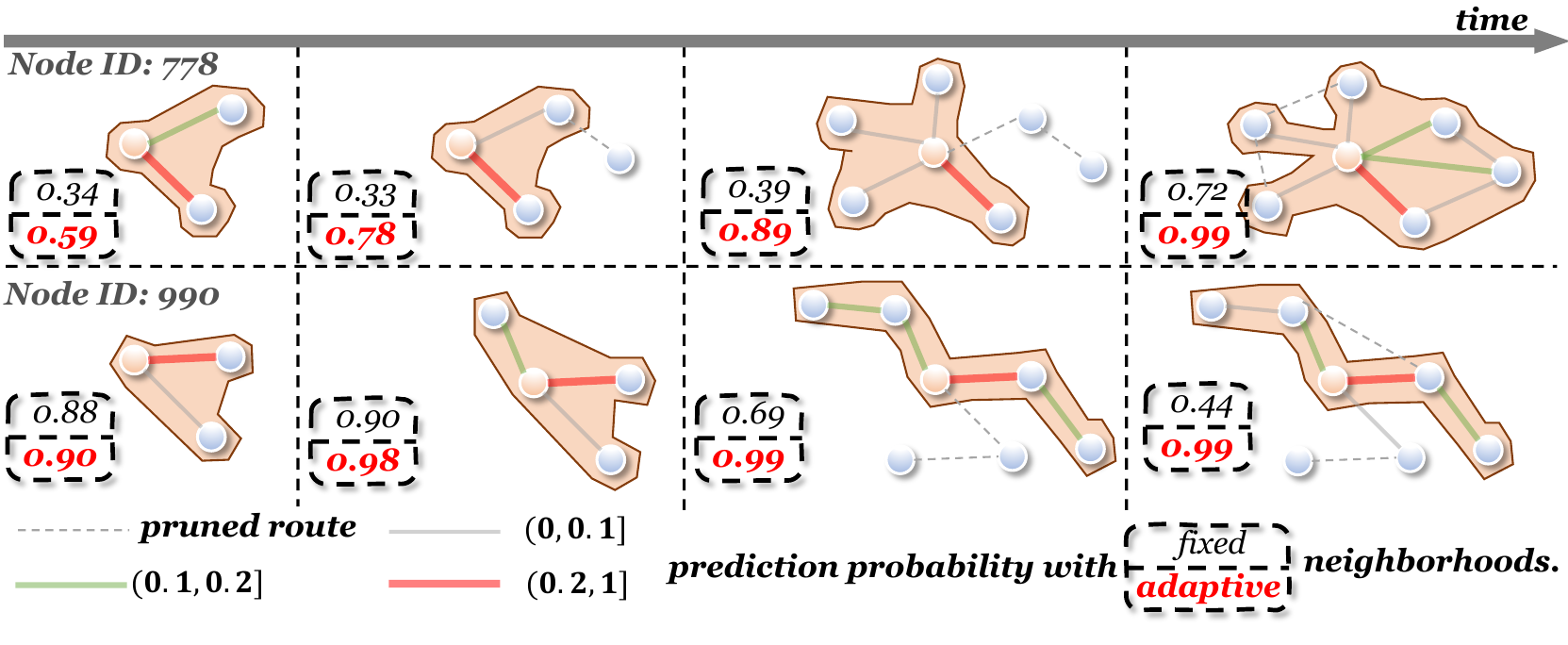}
  \captionsetup{justification=centering}
  \caption{Case study on Wikipedia with TGN model.  \\ (Supplementary results for Section \ref{sec:case})}
  \label{fig:app_casestudy}
\end{figure}

\begin{figure}
    \centering
    \includegraphics[width=\linewidth]{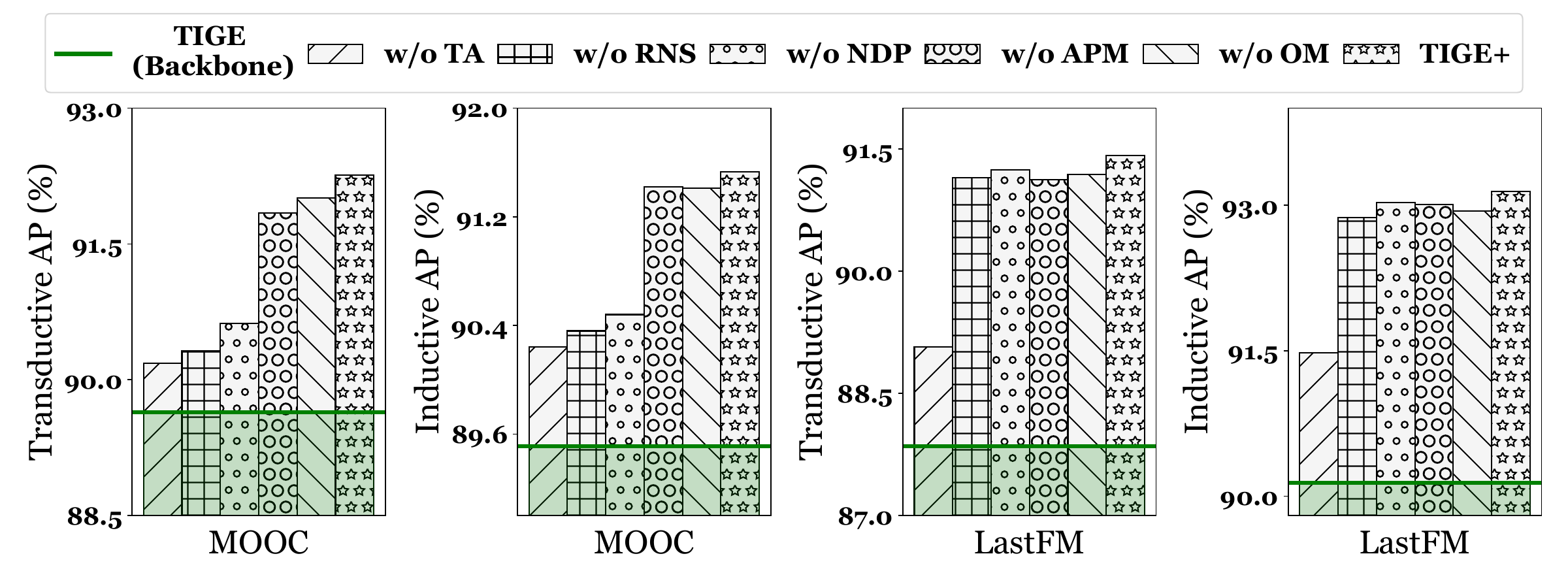}
    \captionsetup{justification=centering}
    \caption{Ablation study on MOOC and LastFM with the improved TIGE model. TIGE backbone is shown for reference.  \\ (Supplementary results for Section \ref{sec:ablation})}
    \label{fig:app_ablation}
\end{figure}

\subsection{Supplementary Results}
Due to space constraints in the main body of our paper, we present the following supplementary results: the robustness study to noisy neighborhoods is depicted in Figure \ref{fig:app_noise_mooc_lastfm} and \ref{fig:app_noise_reddit_score}, the robustness study to expanded neighborhoods is shown in Figure \ref{fig:app_layer}, additional case study with more other selected nodes are illustrated in Figure \ref{fig:app_casestudy}, and the ablation study is detailed in Figure \ref{fig:app_ablation}.

\newpage

(i) Robustness Study to Noisy Neighborhoods.
Supplementary results for noisy MOOC and LastFM are provided. Additionally, we report the box plot of the attention scores for perturbed neighbors on noisy Reddit.
(ii) Robustness Study to Expanded Neighborhoods.
Supplementary results for Wikipedia are presented.
(iii) Case Study.
We present the additional case study derived from other sampled nodes.
(iv) Ablation Study.
Supplementary ablation results for the improved TIGE model and its variants.


\end{document}